\documentclass[prd,nofootinbib,preprint,superscriptaddress]{revtex4-1}

\usepackage[utf8]{inputenc}
\usepackage[T1]{fontenc}
\usepackage[margin=1in]{geometry}
\usepackage{amsmath, amssymb}
\usepackage{graphicx}
\usepackage{hyperref}
\usepackage{cleveref}
\usepackage{tikz}
\usepackage{tikz-3dplot}

\begin{document}

\title{Two Flavon Froggatt-Nielsen Models with Genetic Algorithms}

\author{Miguel Crispim Rom\~ao} 
\email{miguel.romao@durham.ac.uk}
\affiliation{Institute for Particle Physics Phenomenology, Department of Physics, Durham University, Durham DH1 3LE, U.K.}

\author{Stephen F. King}
\email{S.F.King@soton.ac.uk}
\affiliation{School of Physics and Astronomy, University of Southampton,
Southampton SO17 1BJ, United Kingdom}

\begin{abstract}
We present the first systematic and comprehensive scan of two-flavon Froggatt-Nielsen (FN) models, employing artificial intelligence techniques to explore the high-dimensional, mixed discrete--continuous parameter space. Extending the standard single-flavon FN framework to a two-flavon setup in which separate flavon fields couple independently to the up- and down-type sectors, we demonstrate that the relative phase between their vacuum expectation values (vevs) provides a natural and generic source of CP violation absent in single-flavon models.
To explore this enlarged model space, we cast the search for phenomenologically viable models as a multi-objective optimisation problem, formulating each experimental constraint as a separate objective, and employ the Non-dominated Sorting Genetic Algorithm III to simultaneously fit all 18 FN charges, 45 Wilson coefficients, and flavon parameters to both the quark and lepton sectors. Our approach requires no separate training phase and identifies phenomenologically viable models orders of magnitude faster than prior reinforcement learning methods. Imposing experimental constraints on CKM and PMNS mixing angles and CP phases, charged fermion masses, and neutrino squared-mass differences, we discover over $100\,000$ unique viable models with a remarkably low duplication rate, indicating that the space of valid two-flavon FN realisations has not been exhausted. Both Normal and Inverted neutrino mass squared orderings are realised, with the relative hierarchy between the flavon vevs producing qualitatively distinct predictions for the effective neutrinoless double beta decay mass $m_{ee}$. We further demonstrate the existence of minimal FN realisations with maximal flavon exponent as small as three, and of models reproducing charged fermion masses to within $6\%$ without any dedicated continuous parameter optimisation.

\end{abstract}

\maketitle

\tableofcontents

\section{Introduction}

Understanding the origin of the observed pattern of fermion masses and mixing angles is one of the outstanding open problems in particle physics. The Standard Model (SM) accommodates these hierarchies through Yukawa couplings that span several orders of magnitude, yet offers no dynamical explanation for their values. A compelling and widely studied framework for addressing this so-called \emph{flavour puzzle} is the Froggatt--Nielsen (FN) mechanism~\cite{Froggatt:1978nt,Leurer:1993gy}, in which a new Abelian flavour symmetry $U(1)_{FN}$ is introduced alongside one or more scalar \emph{flavon} fields. The observed hierarchies in masses and mixings then arise naturally from the suppression of higher-dimensional operators by powers of the ratio of the flavon vacuum expectation value (vev) to the UV cut-off scale, with the suppression exponent determined by the $U(1)_{FN}$ charge assignments of the matter fields.

Despite being extensively studied in the literature~\cite{Altmannshofer:2024hmr,Cornella:2023zme,Smolkovic:2019jow,Ibe:2024cvi,Cornella:2024jaw}, systematically exploring the model space of FN theories is a challenging task. A FN model is characterised by a set of discrete integer charges for the SM matter fields and one or more continuous parameters governing the flavon vevs and the order-unity Wilson coefficients. Even for the minimal single-flavon case, the search for charge assignments that simultaneously reproduce all measured quark and lepton masses, CKM and PMNS mixing angles, and CP-violating phases constitutes a high-dimensional, highly constrained, and non-linear combinatorial problem. This challenge is compounded in the two-flavon scenario, where the enlarged parameter space and richer operator structure make exhaustive analytical enumeration intractable.

Recent years have witnessed a growing interest in applying artificial intelligence (AI) and machine learning (ML) techniques to model building in high-energy physics (HEP)~\cite{hepmllivingreview}. Methods ranging from neural networks and Bayesian optimisation~\cite{Cranmer:2019eaq,Goodsell:2022beo,Hammad:2022wpq} to reinforcement learning (RL) and genetic algorithms (GAs) have been deployed in diverse contexts, from string landscape surveys~\cite{Abel:2021rrj,Halverson:2019tkf,Cole:2021nnt} to the construction and exploration of beyond-the-Standard-Model (BSM) effective field theories, SUSY and Scotogenic parameter scans,  and multi-Higgs landscapes~\cite{deSouza:2022uhk,Kawai:2024pws,Romao:2024gjx,deSouza:2025uxb,deSouza:2025bpl,Boto:2025mmn,Boto:2025ovp,Boto:2026gzj,deSouza:2026jww,Matchev:2024ash}. In the flavour physics context, reinforcement learning has been applied to FN-type quark mass model building~\cite{Harvey:2021oue} and single-flavon FN model building~\cite{Nishimura:2020nre}, demonstrating that autonomous search methods can identify phenomenologically viable charge assignments without manual enumeration, while more recent work has explored AI-assisted neutrino flavour theory design~\cite{Baretz:2025zsv}. The pioneering study of Ref.~\cite{Nishimura:2020nre}, however, was restricted to the single-flavon case, employed relatively loose experimental constraints, and required a computationally expensive training phase to learn the search policy before viable models could be identified.

In this work, we extend and significantly improve upon this programme in two key respects. First, on the \emph{physics} side, we move beyond the single-flavon framework to the two-flavon scenario, specifically focusing on the class of models in which a flavon $\Phi_u$ couples exclusively to the up-type sector and a flavon $\Phi_d$ to the down-type sector. As we show in~\cref{sec:cpv_two_flavons}, this setup provides a natural and generic source of CP violation through the relative phase between the two flavon vevs~\cite{Kanemura:2007yy}, a feature that is entirely absent in single-flavon models where all flavon-induced phases can be removed by field redefinitions. Second, on the \emph{methodology} side, we replace the RL approach with a multi-objective genetic algorithm, specifically NSGA-III~\cite{deb2013evolutionary,jain2013evolutionary}, in which each experimental constraint is treated as a separate optimisation objective rather than being aggregated into a single scalar loss. This multi-objective formulation allows the algorithm to navigate the tension between different constraints by maintaining a diverse population spread along the Pareto front of the objective space, and avoids the pitfalls of competing loss scales and local minima that plague single-objective formulations. Crucially, unlike RL, our GA-based framework requires no separate training phase and begins identifying viable models immediately, making it far more efficient in practice.

To ensure that the discovered models exhibit genuine FN behaviour -- where Yukawa hierarchies are dynamically generated by flavon suppression rather than tuned through Wilson coefficients -- we impose a hierarchy between the two flavon vevs and restrict the Wilson coefficients to be of order unity. We impose stringent phenomenological constraints: CKM and PMNS angles and CP phases are required to lie within $3\sigma$ of current global fits~\cite{Esteban:2024eli,ParticleDataGroup:2024cfk}, charged fermion masses within $20\%$ of their measured central values, and all flavon exponents are enforced to be non-negative during the search rather than being filtered out \emph{a posteriori}.

This work is organised as follows. In~\cref{sec:flavon_models} we motivate and define the FN model space to explore. Next, in~\cref{sec:scan_methodology} we introduce our scan methodology to explore highly constrained multidimensional model spaces with a multi-objective approach. In~\cref{sec:results} we present the results of our scans, presenting the first thorough two-flavon FN models scan. Finally, in~\cref{sec:conclusions} we conclude. 

\section{Froggatt-Nielsen Models\label{sec:flavon_models}}

For two FN flavons, $\Phi_1$ and $\Phi_2$, the flavon-matter interaction terms Lagrangian read
\begin{align}\label{eq:L_FN_general}
	\mathcal{L} \supset & \sum_{n^u_{1,ij}, n^u_{2,ij}} \left(\frac{\Phi_1}{\Lambda}\right)^{n^u_{1,ij}} \left(\frac{\Phi_2}{\Lambda}\right)^{n^u_{2,ij}} c^u(n^u_{1,ij},n^u_{2,ij}) \bar Q_i u_j H^c   + \text{h.c.}\nonumber\\
    & +\sum_{n^d_{1,ij},n^d_{2,ij}} \left(\frac{\Phi_1}{\Lambda}\right)^{n^d_{1,ij}} \left(\frac{\Phi_2}{\Lambda}\right)^{n^d_{2,ij}} c^d(n^d_{1,ij},n^d_{2,ij}) \bar Q_i d_j H   + \text{h.c.}\textbf{} \nonumber             \\
	& + \sum_{n^\nu_{1,ij}, n^\nu_{2,ij}} \left(\frac{\Phi_1}{\Lambda}\right)^{n^\nu_{1,ij}} \left(\frac{\Phi_2}{\Lambda}\right)^{n^\nu_{2,ij}} c^\nu(n^\nu_{1,ij},n^\nu_{2,ij}) \bar L_i N_j H^c  + \text{h.c.} \nonumber \\
    & + \sum_{n^e_{1,ij},n^e_{2,ij}} \left(\frac{\Phi_1}{\Lambda}\right)^{n^e_{1,ij}} \left(\frac{\Phi_2}{\Lambda}\right)^{n^e_{2,ij}} c^e(n^e_{1,ij},n^e_{2,ij}) \bar L_i e_j H   + \text{h.c.} \nonumber \\
	& + \sum_{n^N_{1,ij},n^N_{2,ij}} \frac{1}{2}  \left(\frac{\Phi_1}{\Lambda}\right)^{n^N_{1,ij}} \left(\frac{\Phi_2}{\Lambda}\right)^{n^N_{2,ij}}\Lambda  c^N(n^N_{1,ij},n^N_{2,ij}) \bar N^c_i N_j \ ,
\end{align}
where $n_{l,ij}^X$ ($X = u, d, \nu, e,  N$, $l = 1,2$) denote the flavon exponents; $Q_i$ ($L_i$) represent the left-handed quark (lepton) $\mathrm{SU}(2)_L$ doublets; $u_i$, $d_i$, and $e_i$ correspond to the right-handed up-type quark, down-type quark, and charged lepton $\mathrm{SU}(2)_L$ singlets, respectively; $N_i$ denote the right-handed neutrino singlets; $H$ is the Higgs doublet field (with $H^c = i\sigma_2 H^*$); $\Lambda$ is the ultraviolet (UV) cut-off scale; and $c^X(n^X_{1,ij},n^X_{2,ij})$ are the Froggatt–Nielsen Wilson coefficients.
In this work, we will focus on the case in which all Wilson coefficients are taken to be real, though not necessarily positive. Under this assumption, these couplings do not induce CP violation, and any CP-violating effects must instead originate from the complex phases of the flavon fields (or, more precisely, from the relative phase between the two flavons).

The sums run over all possible allowed values of $n^X_{l,ij}$, which are constrained by $U(1)_{FN}$ invariance. Denoting $Q(X)$ as the Froggatt-Nielsen charge for the field $X$, the values of $n^X_{l,ij}$ are constrained by the master equations for $U(1)_{FN}$ invariance
\begin{eqnarray}\label{eq:master_equation_general}
	Q(\Phi_1) n^u_{1,ij} + Q(\Phi_2) n^u_{2,ij}     & =  n^u_{ij}    & = - ( Q(\bar Q_i) + Q(u_j) + Q(H^c)  ) \\
	Q(\Phi_1) n^d_{1,ij} + Q(\Phi_2) n^d_{2,ij}     & =  n^d_{ij}    & = - ( Q(\bar Q_i) + Q(d_j) + Q(H)    ) \\
	Q(\Phi_1) n^\nu_{1,ij} + Q(\Phi_2) n^\nu_{2,ij} & =  n^\nu_{ij}  & = - ( Q(\bar L_i) + Q(\nu_j) + Q(H^c)) \\
	Q(\Phi_1) n^e_{1,ij} + Q(\Phi_2) n^e_{2,ij}     & =  n^e_{ij}    & = - ( Q(\bar L_i) + Q(e_j) + Q(H)    ) \\
	Q(\Phi_1) n^N_{1,ij} + Q(\Phi_2) n^N_{2,ij}     & =  n^N_{ij}     & = - ( Q(\bar N^c) + Q(N)) \ ,
\end{eqnarray}
where all equations are subject to the conditions that flavon exponents, $n^X_{1,ij}$ and $n^X_{2,ij}$, are non-negative integers for all $X \in \{u,d,\nu,e,N\}$. 

We first note that the single-flavon scenario can be straightforwardly recovered from the above expressions by setting the exponents associated with one of the flavons to zero in all couplings, i.e. $n^X_{l,ij} = 0$ with $l = 1$ or $2$ for all $X$. If, in addition, we impose $|Q(\Phi_l)| = 1$, where $l$ now labels the only active flavon field, we reproduce the one-flavon model studied in~\cite{Nishimura:2020nre}. As discussed in~\cref{sec:cpv_single_flavon}, the single-flavon case does not provide an explanation for the origin of CP violation, even when the flavon vev carries a non-zero complex phase.

Focusing on the two-flavon cases, an important observation is that the number of FN operators contributing to the same Yukawa coupling depends on the FN charges of the Higgs and matter fields. In particular, the sums can be infinite if the flavons charges have opposite signs. To see this, consider $Q(\Phi_1)>0$ and $Q(\Phi_2)<0$. In this case, the master condition reads
\begin{equation}
	Q(\Phi_1)\,n^X_{1,ij} - |Q(\Phi_2)|\,n^X_{2,ij} =  n^X_{ij}\ ,
\end{equation}
which can be solved for one of the flavon exponents as
\begin{equation}
	n^X_{1,ij}  = \frac{1}{Q(\Phi_1)}\bigl(  n^X_{ij} + |Q(\Phi_2)|\,n^X_{2,ij} \bigr) \geq 0 \ ,
\end{equation}
implying the constraint
\begin{equation}
	 n^X_{ij} + |Q(\Phi_2)|\,n^X_{2,ij} \geq 0 \quad  \Leftrightarrow \quad n^X_{2,ij} \geq - \frac{ n^X_{ij}}{|Q(\Phi_2)|}\,.
\end{equation}
This condition enforces $n^X_{ij} \leq 0$, but it does not yield an upper bound on $n^X_{2,ij}$, which remains restricted only to be a non-negative integer. Consequently, this configuration corresponds to an infinite sum: the opposite signs of the flavon FN charges permit cancellations on the left-hand side of~\cref{eq:master_equation_general}, leaving $n^X_{ij}$ without a finite upper constraint.

Conversely, if the flavon charges have the same sign, we can find an upper bound on the sums, i.e. there is a finite number of FN operators. To see this, consider $Q(\Phi_1)>0$, $Q(\Phi_2)>0$. In this case, the exponents satisfy
\begin{equation}
	Q(\Phi_1)\,n^X_{1,ij} + Q(\Phi_2)\,n^X_{2,ij} = n^X_{ij}\,,
\end{equation}
which can be solved for one of the flavon exponents as
\begin{equation}
	n^X_{1,ij} = \frac{1}{Q(\Phi_1)} \bigl( n^X_{ij} - Q(\Phi_2)\,n^X_{2,ij} \bigr) \geq 0\,.
\end{equation}
This condition implies
\begin{equation}
	n^X_{ij} - Q(\Phi_2)\,n^X_{2,ij} \geq 0 
	\;\Leftrightarrow\;
	n^X_{2,ij} \leq \frac{n^X_{ij}}{Q(\Phi_2)}\,,
\end{equation}
which is only valid for $n^X_{ij} \geq 0$, and in that regime it provides an upper bound on the allowed values in the summations over $n^X_{1,ij}$ and $n^X_{2,ij}$. While providing a finite number of FN operators, this bound still allows for its number to depend explicitly on the FN charge assignments, which in turn set $n^X_{ij}$. This poses a great challenge to systematically study the two-flavon cases as the parameter space does not have a fixed dimension across all possible FN charge assignments. Instead, we will focus on an UV-motivated scenario where the two flavons couple separately to the up and down sectors.

\subsection{Up and Down-type Flavon Models}

We now consider the scenario in which each flavon couples independently and exclusively to either the up- or down-type sectors, i.e. we define $\Phi_1 = \Phi_u$ and $\Phi_2 = \Phi_d$, where $\Phi_u$
couples only to up-type quarks and 
$\Phi_d$
couples only to down-type quarks,
and set
\begin{align}
    n^u_{d,ij} = n^\nu_{d,ij} = n^N_{d,ij} = n^d_{u,ij} = n^e_{u,ij} = 0  \ .
\end{align}
This setup is motivated by UV completions in which the up-type and down-type Yukawa couplings are generated by distinct interactions
(see e.g.~\cite{King:2003rf}).
In this case, the Lagrangian takes the form
\begin{align}\label{eq:L_FN_ud}
	\mathcal{L} \supset & \sum_{n^u_{u,ij}} \left(\frac{\Phi_u}{\Lambda}\right)^{n^u_{u,ij}}  c^u(n^u_{u,ij}) \,\bar Q_i u_j H^c 
	+ \sum_{n^d_{d,ij}}\left(\frac{\Phi_d}{\Lambda}\right)^{n^d_{d,ij}}  c^d(n^d_{u,ij})  \bar Q_i d_j H 
	+ \text{ h.c.} \nonumber       \\
	& + \sum_{n^\nu_{u,ij}} \left(\frac{\Phi_u}{\Lambda}\right)^{n^\nu_{u,ij}}  c^\nu(n^\nu_{u,ij}) \,\bar L_i N_j H^c 
	+ \sum_{n^e_{d,ij}}  \left(\frac{\Phi_d}{\Lambda}\right)^{n^e_{d,ij}}  c^e(n^e_{u,ij}) \bar L_i e_j H 
	+ \text{ h.c.} \nonumber \\
	& + \sum_{n^N_{u,ij}} \frac{1}{2}  c^N(n^N_{u,ij}) \left(\frac{\Phi_u}{\Lambda}\right)^{n^N_{u,ij}}  \Lambda \,\bar N^c_i N_j \,,
\end{align}
where we further assume that the right-handed neutrino effective Majorana mass matrix is solely generated by the up-type flavon. This Lagrangian leads to the simplified master relations
\begin{align}\label{eq:up-down-master-equations}
	n^X_{u/d,ij} =  \frac{n^X_{ij}}{Q(\Phi_{u/d})} \,,
\end{align}
where $X = u, d, \nu, e, N$ and $Q(\Phi_{u/d})$ denotes the corresponding $U(1)_{FN}$ charge of the flavon. Since $n^X_{u/d,ij}$ must be non-negative, it follows that $n^X_{ij}$ must be positive (negative) if $Q(\Phi_{u/d})$ is positive (negative). These constraints can be interpreted as Kronecker-delta–like selection rules, which trivialise the sums, yielding
\begin{align}\label{eq:L_FN_ud_simplified}
	\mathcal{L}  \supset & \left(\frac{\Phi_u}{\Lambda}\right)^{ n^u_{ij}/Q(\Phi_u)}  c^u_{ij} \,\bar Q_i u_j H^c 
	+ \left(\frac{\Phi_d}{\Lambda}\right)^{ n^d_{ij}/Q(\Phi_d)} c^d_{ij} \,\bar Q_i d_j H 
	+ \text{ h.c.} \nonumber      \\
	& + \left(\frac{\Phi_u}{\Lambda}\right)^{ n^\nu_{ij}/Q(\Phi_u)} c^\nu_{ij} \,\bar L_i N_j H^c 
	+ \left(\frac{\Phi_d}{\Lambda}\right)^{ n^e_{ij}/Q(\Phi_d)} c^e_{ij} \,\bar L_i e_j H 
	+ \text{ h.c.} \nonumber \\
	& + \frac{1}{2} c^N_{ij} \left(\frac{\Phi_u}{\Lambda}\right)^{n^N_{ij}/Q(\Phi_u)} \Lambda \,\bar N^c_i N_j \,,
\end{align}
where $c^X_{ij}=c^X(n^X_{l,ij}= n^X_{ij}/Q(\Phi_l))$ are now a single matrix instead of a sum of many. This Lagrangian closely resembles the single-flavon case, in which no explicit sums appear, and each Yukawa coupling is generated by a single  FN Wilson coefficient, greatly reducing the parametric freedom of the model. Therefore, in this case, the flavon charges need not to have the same sign to ensure a finite number of terms. Nonetheless, in this work we will set $Q(\Phi_u)=Q(\Phi_d)=1$ to simplify the handling of the master expressions.\footnote{This is not the most general case. The most general configuration allows non-vanishing FN charges for the matter fields smaller than the flavon charges. However, this requires dealing with non-integer $n^X_{ij}$, while $n^X_{l,ij}$ remain integers. In other words,~\cref{eq:up-down-master-equations} would impose additional Diophantine constraints on our search. If the flavon charges have magnitudes larger than unity, both charges remain free parameters that must be constrained to satisfy the master equation. Nonetheless, it is well motivated to assume that all FN charges can be normalised to integers, with at least one field carrying unit charge. This assumption is justified if $U(1)_{FN}$ is embedded in a larger (semi-)simple UV group. In that case, the corresponding charges share a common irrational factor, which can be factored out and absorbed into the Lagrangian couplings.}

With the setup describe above, we now parametrise the flavon vevs that break $U(1)_{FN}$ as $\epsilon_u = \langle \Phi_u\rangle/\Lambda$ and $\epsilon_d = \langle H_d\rangle/\Lambda$. We also define the relative phase between the vevs as $\epsilon_u/\epsilon_d = e^{i \alpha}$ and, w.l.o.g., we perform an $U(1)_{FN}$ rotation such that $\epsilon_u = | \epsilon_u| e^{i\alpha}$, $\epsilon_d = |\epsilon_d|$. With this parametrisation choice, the Lagrangian is given by
\begin{align}\label{eq:L_FN_ud_parametrised}
	\mathcal{L} \supset &\; \epsilon_u^{n^u_{ij}}\, c^u_{ij}\, \bar Q_i u_j H^c 
	+ \epsilon_d^{n^d_{ij}}\, c^d_{ij} \bar Q_i d_j H + \text{ h.c.} \nonumber \\
	& + \epsilon_u^{n^\nu_{ij}}\, c^\nu_{ij}\, \bar L_i N_j H^c 
	+ \epsilon_d^{n^e_{ij}}\, c^e_{ij} \bar L_i e_j H + \text{ h.c.} \nonumber \\
	& + \frac{1}{2}\, \epsilon_u^{n^N_{ij}} c^N_{ij}\, \,\Lambda\, \bar N^c_i N_j \, .
\end{align}
 From this, we can immediately read the Yukawa couplings and the effective Majorana mass term for the right handed neutrino,
\begin{align}
    Y^u_{ij} &= \epsilon_u^{n^u_{ij}}\, c^u_{ij} , \, Y^d_{ij} = \epsilon_d^{n^d_{ij}} c^d_{ij}, \nonumber\\
    Y^\nu_{ij} &= \epsilon_u^{n^\nu_{ij}}\, c^\nu_{ij} , \, Y^e_{ij} = \epsilon_d^{n^e_{ij}} c^e_{ij},\nonumber\\
    M^N_{ij}  &= \epsilon_u^{n^N_{ij}} c^N_{ij}\, \,\Lambda \, .
\end{align}
A key feature of this class of models is that the relative phase $e^{i\alpha}$ between the flavon vacuum expectation values cannot be removed by field redefinitions. Consequently, one generically expects CP violation, as discussed in more detail in~\cref{sec:cpv_two_flavons}.
In order to have a proper FN model, where the flavour structure is being set by the FN coefficients and flavon suppression instead of the Wilson coefficients, we need $|\epsilon_u|, |\epsilon_d| \ll 1$. Furthermore, we will want to study the cases where one of the flavon vevs clearly leads over the other.
In the analysis conducted below, we will treat these two configurations separately during the parameter scan, but the outcomes will be combined when presenting the final results.

\subsection{Observable Conventions and Definitions}

We briefly define our conventions that link a model prediction to an observable. The effective Yukawa couplings below FN spontaneous symmetry breaking (SSB) is
\begin{equation}
\mathcal{L}_m \supset \bar L_i (Y^e H)_{ij} e_j + \bar L_i (Y^\nu H^c)_{ij} N_j + \bar Q_i (Y^d H)_{ij} d_j + \bar Q_i (Y^u H^c)_{ij} u_j  + \frac{1}{2} \bar N^c_i (M^N)_{ij} N_j   + \text{h.c.}
\end{equation}
which, below the electroweak SSB leads to the Dirac masses
\begin{equation}
    \mathcal{L}_m \supset \bar E_i (m^e)_{ij} e_j + \bar \nu_i (m^\nu_D)_{ij} N_j + \bar D_i (m^d)_{ij} d_j + \bar U_i (m^u)_{ij} u_j  + \frac{1}{2} \bar N^c_i (M^N)_{ij} N_j   + \text{h.c.} \ ,
\end{equation}
and the neutrino masses are generated via a type-I see-saw mechanism
\begin{equation} 
    M_\nu = - m^\nu_D (M^N)^{-1} (m^\nu_D)^T \ .
\end{equation}

For the charged leptons, their mass matrices, $m^X$, can be rotated into a diagonal form, $\Delta^X$, using two unitary matrices, $U^X$ and $V^X$, 
\begin{equation}
    (m^X)_{ij} = (U^X)_{iI} \Delta^X_{IJ} (V^X)^\dagger_{Jj}  \\
\Leftrightarrow (U^X)^\dagger_{Ii}  (m^X)_{ij} (V^X)_{jJ} =  \Delta^X_{IJ} 
\end{equation}
where here, and in what follows, $i,j,...$ are flavour basis indices and $I,J,...$ are mass basis indices. The $U^X$ ($V^X$) matrices can then seen as rotating the left (right) handed fields, such that
\begin{align}
    \tilde X_L = (U^X)^\dagger X_L \\
\tilde X_R = (V^X)^\dagger X_R \ , 
\end{align}
where the tildes refer to the mass basis. With these definitions the charged fermion fields and masses are rotated to the diagonal mass basis as
\begin{equation}
    (\bar X_L)_i m^X_{ij} (X_R)_j = (\bar X_L)_i (U^X)_{iJ} (U^X)^\dagger_{Jk} m^X_{kl} (V^X)_{lL} (V^X)^\dagger_{Lk} (X_R)_k  \\
= (\bar{\tilde{X}}_L)_I \Delta^X_{IJ}(\tilde{X}_R)_{J} \ .
\end{equation}

The neutrinos have a slightly different rotation prescription as their mass matrix is complex symmetric. In this case, one can rotate to the diagonal mass basis using a complex orthogonal matrix $U^\nu$, such that
\begin{equation}
    (M_\nu)_{ij} = (U^\nu)_{iI} \Delta^\nu_{IJ} (U^\nu)^T_{Jj} \ ,
\end{equation}
or, conversely,
\begin{equation}
    (U^\nu)^\dagger_{Ii} (M_\nu)_{ij} (U^\nu)^*_{jJ} =  \Delta^\nu_{IJ}  \ .
\end{equation}

With these definitions one then arrives at our convention for the CKM and PMNS matrices
\begin{align}
    U_{CKM} &= (U^u)^\dagger U^d \\
U_{PMNS}    &= (U^e)^\dagger U^\nu \ .
\end{align}

\subsubsection{PMNS \texttt{NuFit} Convention}

To connect the PMNS and neutrino mass predictions to data we need to further align our conventions with those used by the~\texttt{NuFit} collaboration~\cite{Esteban:2024eli}, which communicates the leading global fits to neutrino data. To this effect, we first make use of the generic $3\times3$ unitary matrix parametrisation, which is often used to express for example the CKM, as
\begin{equation}
    \bar{U}(\theta_{12},\theta_{13},\theta_{23},\delta) = \begin{pmatrix}
c_{12} c_{13} & s_{12} c_{13} & s_{13} e^{-i\delta} \\
-s_{12} c_{23} - c_{12} s_{23} s_{13} e^{i\delta} & c_{12} c_{23} - s_{12} s_{23} s_{13} e^{i\delta} & s_{23} c_{13} \\
s_{12} s_{23} - c_{12} c_{23} s_{13} e^{i\delta} & -c_{12} s_{23} - s_{12} c_{23} s_{13} e^{i\delta} & c_{23} c_{13}
\end{pmatrix}
\end{equation}
where $s_{ij}$ ($c_{ij}$) are the $\sin$ ($\cos$) of the rotation angles on the $ij$ plane, i.e. of the three Euler angles, and $\delta$ is the CP-violating phase that cannot be rotated away. This matrix can then be decomposed as
\begin{equation}
    \bar{U}(\theta_{12},\theta_{13},\theta_{23},\delta) = R_{23}(\theta_{23}) \cdot \Gamma(\delta) \cdot R_{13}(\theta_{13}) \cdot \Gamma^\dagger(\delta) \cdot R_{12}(\theta_{12}) \ ,
\end{equation}
where $R_{ij}$ are $SO(3)$ rotations on the $ij$ planes, and
\begin{equation}
    \Gamma(\delta) = \text{diag}(1, 1, e^{i\delta}) \ .
\end{equation}

With these definitions, the PMNS matrix can then be defined as having a further diagonal phase contribution that can be factored to the right following the~\texttt{NuFit} convention, i.e.
\begin{equation}
    U_{PMNS} = \bar U \cdot P \ ,
\end{equation}
with
\begin{equation}
    P  = \text{diag}(e^{i\alpha_1}, e^{i\alpha_2}, 1) \ ,
\end{equation}
which then set our definition for the two Majorana phases $\alpha_1$ and $\alpha_2$.

The above conventions allow us to compare PMNS predictions to data. For the neutrino squared mass differences, we first need correctly identify the ordering. To do so, we first work in the basis where the charged leptons are diagonalised, i.e.
\begin{equation}
    M_\nu^{\text{e-flavour}} = (U^{e})^\dagger M_\nu (U^{e})^* \ ,
\end{equation}
which, in this basis, is diagonalised by $\tilde U^\nu$, and the corresponding neutrino mass eigenstates are labelled $\nu_1, \nu_2, \nu_3$ in agreement with the~\texttt{NuFit} convention:
\begin{itemize}
    \item Identify the solar pair: $\nu_1$ and $\nu_2$ are the two states with the largest electron flavour content, $|\tilde U^\nu_{ei}|$.
    \item Set mass hierarchy within solar pair: $m_2 > m_1$, which ensures $\Delta m_{21}^2 > 0$.
    \item Identify the ordering:
    \begin{itemize}
    \item Normal Ordering (NO): $m_3 > m_2 > m_1$ \ ,
    \item  Inverted Ordering (IO): $m_2 > m_1 > m_3$ \ .
    \end{itemize}
\end{itemize}
And the ``final'' PMNS matrix is $\tilde U^\nu$ with the indices $1,2,3$ ordered as above.

\section{Scan Strategy and Methodology\label{sec:scan_methodology}}

The main challenge in constructing FN models is exploring a high-dimensional, highly constrained, and non-linear parameter space. A model is defined by discrete integer charges for the matter fields, continuous Wilson coefficients, and the magnitudes and phases of the flavon vevs. We search for parameter configurations $\mathbf{x}$ that satisfy physical constraints on $n_O$ observables $O_k(\mathbf{x})$, with $k = 1, \dots, n_O$, whose allowed ranges are fixed by experimental data such as fermion masses and mixing parameters. Our goal is to systematically scan this space to find regions where $\mathbf{x}$ yields phenomenologically viable FN models. To do so, we use a Genetic Algorithm (GA), well-suited to navigating such complex and diverse spaces, as discussed below.

\subsection{Comparison with Reinforcement Learning Approaches}

A related FN model-building strategy was recently explored in~\cite{Nishimura:2020nre} using Reinforcement Learning (RL).\footnote{See also~\cite{Harvey:2021oue,Baretz:2025zsv} for RL applications to model building.} There, an RL agent scanned $U(1)_{FN}$ flavour charges to find viable single-flavon models in a two-stage search: first fitting the quark sector FN charges, then the lepton sector FN charges, while optimising for the Wilson the flavon vev separately. While promising, the RL paradigm differs conceptually and practically from the GA framework used in this work, particularly in training cost, applicability, and model-discovery features.

RL methods typically require a potentially expensive training phase to learn an effective search policy before systematic parameter-space exploration, and involve many hard-to-tune hyperparameters. In contrast, GAs such as the one we use and will be described below, are direct optimisation methods without a separate training stage, beginning exploration immediately. This difference grows more important as the optimisation dimensionality increases. Our framework efficiently explores a much larger space -- simultaneously fitting all 18 FN charges, 45 Wilson coefficients, and flavon vev parameters, to both quark and lepton sectors (including CP-violating phases) -- without any intermediate search policy-learning step.

These methodological differences directly affect model-discovery efficiency. In an initial stage of our work, we observed that in the single-flavon scenario, our GA-based approach finds thousands of distinct models satisfying all phenomenological constraints (jointly fitting quark and lepton data) within minutes, compared to the $\mathcal{O}(10)$ models of~\cite{Nishimura:2020nre}, obtained after about reported fifteen hours of RL training on the quark sector alone. This efficiency gain stems from directly exploring the solution space and formulating the task as a multi-objective optimisation problem, enabling a systematic, parallel search for mutually distinct yet phenomenologically promising models. In this phenomenological context, the main scientific interest lies in the set of viable models (the optimisation solutions), not in the decision-making search policy whose learning dominates the RL computational and runtime costs. GAs are intrinsically solution-centric, making them a natural and efficient tool for charting viable FN models.

This is not to say that RL lacks advantages. A key benefit over GAs is that task-optimisation information is explicitly encoded in the neural network parametrising the search policy, whereas in a GA it is only implicitly contained in the evaluated points, without an explicit learned mapping of the model landscape. Thus, given enough compute and training time, RL may achieve superior asymptotic exploration. Ref.~\cite{Abel:2021rrj} systematically compared GA and RL in exploring Standard Model–like configurations in the heterotic string landscape. In that setting, GA was about an order of magnitude more efficient: the RL-based scan needed at least an order of magnitude longer runtime to find the same number of viable configurations. A broader, systematic comparison between RL and GA for BSM model building is left for future work.

\subsection{Constraint Losses and Multi-Objective Optimisation}

To identify phenomenologically viable models, we formulate the search as a multi-objective optimisation problem. This enables efficient scans with a wide exploration. We use a loss function to encode each of the relevant constraints, with each acceptance criteria  tailored to the precision requirements of this analysis and to the theoretical uncertainties of each class of observables.

\subsubsection{Constraint Losses}

Our goal is to find models compatible with experimental flavour data -- fermion masses, mixing angles, and CP-violating phases -- which have very different uncertainties. For example, the electron mass is known with extremely high precision, whereas the neutrino CP-violating phase is still poorly constrained. Moreover, fermion masses are usually given as pole masses, so a fully consistent treatment would require running them to a common reference scale via a detailed renormalisation group analysis.

Such sophistication exceeds the intended precision of this work, whose main goal is to find potentially viable FN models rather than to deliver a detailed phenomenological study of any specific model. We therefore handle different observables with different levels of precision requirements: For PMNS and CKM quantities and neutrino square mass differences, we use their experimental uncertainties to define a quantitative criterion for phenomenological viability, while for the charged fermion masses we set a common relative uncertainty interval.

To encode the deviation from the acceptance criteria, we adopt a ``$C$-type'' loss, introduced in Ref.~\cite{deSouza:2022uhk} and successfully applied in BSM parameter-space scans in Refs.~\cite{Romao:2024gjx,deSouza:2025uxb,deSouza:2025bpl,Boto:2025mmn,Boto:2025ovp,Boto:2026gzj}. It is defined as
\begin{equation}
    \mathcal{L}_{k} = C(O_k, L_k, U_k) = \max\bigl(0, L_k - O_k(\mathbf{x}), O_k(\mathbf{x}) - U_k\bigr) \,,
\end{equation}
where $L_k$ and $U_k$ typically coincide with the lower and upper bound of the allowed values, often taken to be within a certain significance level but can be arbitrarily set.\footnote{The same construction can also be applied to non-experimental constraints, such as theoretical consistency or to enforce desired features in the solutions.} This piecewise-linear function vanishes when the observable lies within the chosen interval and increases linearly outside it, quantifying the deviation from the allowed range.

The full list of constraints imposed in our work are as follow:
\begin{itemize}
    \item PMNS angles and CP violating phase, and neutrino square mass differences within $3\sigma$ of the \texttt{NuFit} collaboration~\cite{Esteban:2024eli} fit, as presented in~\cref{tab:neutrino_data}.
    \item CKM angles and CP violating phase within $3\sigma$ of experimental data reported in the Particle Data Group~\cite{ParticleDataGroup:2024cfk}, as presented in~\cref{tab:ckm_data}.
    \item Charged fermion masses within within $20\%$ central value as reported in the Particle Data Group~\cite{ParticleDataGroup:2024cfk}, as presented in~\cref{tab:masses_central_values}.
    \item Total number of negative flavon exponents to be strictly zero.
\end{itemize}

\begin{table}[]
    \centering
    \begin{tabular}{cc}
        \hline\hline
         $\theta^{PMNS}_{12} = (33.68\pm0.73)$ & $\Delta m^2_{12}=(7.49\pm0.19)\times10^{-5}$ eV (NO and IO)   \\
         $\theta^{PMNS}_{13} = (8.52\pm0.11)$  & $\Delta m^2_{3l}=(2.534\pm0.025)\times10^{-3}$ eV (NO)  \\
         $\theta^{PMNS}_{23} = ( 48.5\pm0.9)$  & $\Delta m^2_{3l}=(-2.510\pm0.025)\times10^{-3}$ eV (IO)  \\
         $\delta^{PMNS} = (177\pm 20)$         & \\ \hline\hline
    \end{tabular}
    \caption{Neutrino data, angles in degrees. NO (IO) refers to Normal (Inverted) Ordering, which will be detailed further below. Values from the \texttt{NuFit} collaboration~\cite{Esteban:2024eli}.}
    \label{tab:neutrino_data}
\end{table}

\begin{table}[]
    \centering
    \begin{tabular}{cc}
        \hline\hline
         $\sin \theta^{CKM}_{12} = 0.22501\pm0.00068$ & $\sin \theta^{CKM}_{23} = 0.04183\pm0.00079$   \\
         $\sin \theta^{CKM}_{13} = 003732\pm0.000090$  & $\delta^{CKM} = 1.147\pm 0.026$ \\
 \hline\hline
    \end{tabular}
    \caption{DKM data, CP phase in radians. Values from Particle Data Group~\cite{ParticleDataGroup:2024cfk}.}
    \label{tab:ckm_data}
\end{table}

\begin{table}[h]
    \centering
    \begin{tabular}{ccc}
    \hline
    \hline
         $m_u/$ MeV & $m_c/$ GeV & $m_t/$ GeV  \\
         $2.16$     & $1.273$   & 172.56 \\ \hline
         $ m_d/$ MeV & $m_s/$ MeV & $m_b/$ GeV\\
         $4.70$ & 93.5 & $4.183$ \\\hline
         $m_e/$ MeV & $m_\mu/$ GeV & $m_\tau/$ GeV  \\
         $0.511$     & $0.106$   & 1.777 \\ \hline\hline
    \end{tabular}
    \caption{Central values of the charged lepton masses. Values from Particle Data~\cite{ParticleDataGroup:2024cfk}.}
    \label{tab:masses_central_values}
\end{table}

These constraints are far more stringent than those considered in~\cite{Nishimura:2020nre}, where the masses were allowed to vary within a full order of magnitude around the central value, the angles and phases to vary up to $~58.5\%$ away from the central value, and no negative flavon exponents constraints was added during the search, with negative flavon exponents manually removed after scan.

\subsubsection{Combined Multi-objective and Pareto Front}

A point in parameter space is a ``good point'' (i.e. a valid model) if all $C$-type losses vanish, $\mathcal{L}_k(\mathbf{x}) = 0$. There are 17 constraint losses in total: four for the CKM mixing angles and CP phase, four for the PMNS angles and missing phase, three for the quark masses, three for charged-lepton masses, two for the neutrino square mass differences, and one for the number of negative flavon exponents. One could aggregate all constraints into a single scalar loss, as in~\cite{Nishimura:2020nre,deSouza:2022uhk,Romao:2024gjx,deSouza:2025bpl,Boto:2025mmn,Boto:2025ovp,Boto:2026gzj}, but this would sum contributions differing by several orders of magnitude, typically hindering convergence and increasing the risk of trapping in local minima.

An alternative strategy is to use a genuinely multi-objective optimisation paradigm, in which all constraints are optimised \emph{simultaneously} with no \emph{a priori} preference among them, as first done in~\cite{deSouza:2025uxb} when exploring the highly constrained multi-dimensional parameter space of a Scotogenic model. Here, the optimisation target is the vector of all constraints in the \emph{objective space}, which in our case is $\mathbb{R}^{17}$, and can be written as
\begin{equation}\label{eq:objective_vector}
    (\mathcal{L}_1(\mathbf{x}),\dots,\mathcal{L}_{17}(\mathbf{x})) \ \ ,
\end{equation}
commonly called the \emph{objective vector}.

In this space, all phenomenologically viable models lie at the origin, where all constraints are satisfied. This suggests defining a ``promising'' point as one that is ``closest'' to the origin. However, because the constraints have different scales, this distance is of limited use and suffers from the same issues as combining multiple loss terms into a single scalar objective. A more suitable notion of ``promising'' solutions is given by the \emph{Pareto front}, which we now briefly introduce.

Consider the multi-objective minimisation problem with objective vector~\cref{eq:objective_vector} and decision variables $\mathbf{x}$. For two candidate solutions $\mathbf{x}_1, \mathbf{x}_2$, we say that $\mathbf{x}_1$ \emph{dominates} $\mathbf{x}_2$, $\mathbf{x}_1 \prec \mathbf{x}_2$, if
\begin{equation}
  \mathbf{x}_1 \prec \mathbf{x}_2
  \iff
  \begin{cases}
    \mathcal{L}_i(\mathbf{x}_1) \le \mathcal{L}_i(\mathbf{x}_2), & \forall i = 1, \dots, n_O, \\
    \mathcal{L}_j(\mathbf{x}_1) < \mathcal{L}_j(\mathbf{x}_2), & \text{for at least one } j.
  \end{cases}
\end{equation}
In other words, $\mathbf{x}_1$ is strictly better than $\mathbf{x}_2$ in at least one objective and no worse in all others, which defines a partial order over candidate solutions and a notion of multi-objective optimality.

A point $\mathbf{x}^*$ is \emph{Pareto optimal} if no other candidate solution dominates it, i.e.
\begin{equation}
  \mathbf{x}^* \text{ is Pareto optimal}
  \iff 
  \nexists \mathbf{x} : \mathbf{x} \prec \mathbf{x}^*.
\end{equation}

\begin{figure}[h]
    \centering
    \begin{tikzpicture}[>=stealth, scale=1.3]
        \draw[->] (0,0) -- (4.5,0) node[anchor=west] {$\mathcal{L}_1(\mathbf{x})$};
        \draw[->] (0,0) -- (0,3.5) node[anchor=south] {$\mathcal{L}_2(\mathbf{x})$};

        \filldraw[black] (0,0) circle (1.2pt) node[anchor=north east] {$\mathbf{0}$};

        \draw[thick,red]
            (0.5,3.0) .. controls (1.0,2.2) and (2.0,1.4) ..
            (3.5,0.7);

        \filldraw[red] (0.8,2.6) circle (1.6pt) node[anchor=south west] {$\mathbf{x}_{1}$};
        \filldraw[red] (1.725, 1.725) circle (1.6pt) node[anchor=north east] {$\mathbf{x}_{2}$};
        \filldraw[red] (3.1, 0.9) circle (1.6pt) node[anchor=south west] {$\mathbf{x}_{3}$};

        \filldraw[blue] (3.2,2.6) circle (1.6pt) node[anchor=south west] {$\mathbf{y}$};
        \filldraw[blue] (2.25,2.7) circle (1.6pt) node[anchor=south west] {};
        \filldraw[blue] (1.7,2.25) circle (1.6pt) node[anchor=south west] {};
        \filldraw[blue] (2.5, 1.6) circle (1.6pt) node[anchor=south west] {};
        \filldraw[blue] (3.2, 2.1) circle (1.6pt) node[anchor=south west] {};

        \draw[->,blue!70!black,thick] (3.2,2.6) -- (1.8,1.8);

    \end{tikzpicture}
    \caption{Schematic illustration of a Pareto front in a two-dimensional objective space spanned by the losses $\mathcal{L}_1(\mathbf{x})$ and $\mathcal{L}_2(\mathbf{x})$. The red curve represents the set of non-dominated points, i.e.\ the Pareto front $\mathcal{P}$, whose elements $(\mathcal{L}_1(\mathbf{x}_{a}),\mathcal{L}_2(\mathbf{x}_{a}))$ cannot be improved in one loss without worsening at least one other. All blue points are domianted by at least one point in the Pareto front, e.g. the point $(\mathcal{L}_1(\mathbf{y}),\mathcal{L}_2(\mathbf{y}))$ is dominated, since there exists a point on the front, such as $(\mathcal{L}_1(\mathbf{x}_{2}),\mathcal{L}_2(\mathbf{x}_{2}))$, that is no worse in either loss and strictly better in at least one.}
    \label{fig:pareto_front}
\end{figure}
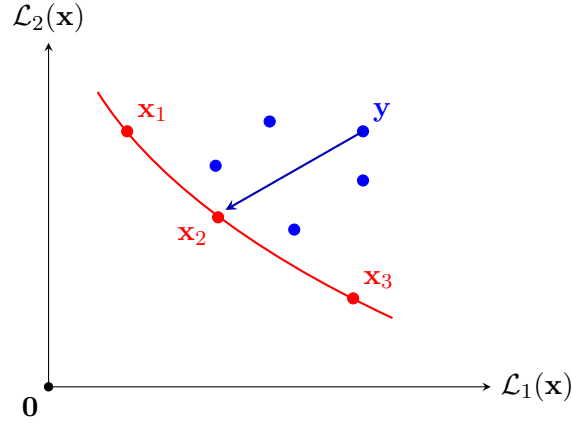

Since dominance requires a candidate solution to be strictly better in at least one objective and no worse in all others, multiple Pareto optimal points can coexist. This leads to the definition of the \emph{Pareto front}, the set of objective vectors of all Pareto optimal solutions:
\begin{equation}
    \mathcal{P} = \left\{ \mathcal{L}_i(\mathbf{x}) : \mathbf{x} \text{ is Pareto optimal} \right\}.
\end{equation}
\Cref{fig:pareto_front} shows a two-dimensional schematic of the Pareto front in objective space. The red points dominate all blue points but not one another; each blue point is dominated by at least one point on the Pareto front.

On the basis of these definitions, one can develop an optimisation paradigm that does not aim at minimising a single scalar loss, but instead iteratively generates new solutions that dominate those previously obtained, thereby updating the Pareto front at each iteration and pushing it toward the origin of the objective space, where all constraints are satisfied.

\subsection{NSGA-III: Diversity in High-Dimensional Spaces}

For the multi-objective optimisation we use the Non-dominated Sorting Genetic Algorithm III (NSGA-III)~\cite{deb2013evolutionary,jain2013evolutionary}, widely regarded as the state-of-the-art GA for multi-objective problems with high-dimensional objective spaces. NSGA-III maintains population diversity via a structured set of reference directions, as we briefly explain here, which enhances exploration in high dimensional spaces. Treating constraints as separate objectives allows the algorithm to handle tensions between observables -- where improving one worsens another -- by distributing the population along the Pareto front rather than collapsing to a single local minimum. NSGA-III has already been successfully applied to the highly constrained, multi-dimensional parameter space of a Scotogenic model in~\cite{deSouza:2025uxb}, yielding a diverse set of viable parameters.

NSGA-III is a GA that operates on a population $P$ of $N$ candidate solutions, $\{\mathbf{x}_i\}_{i=1}^{N}$, each encoding a configuration of FN charges, Wilson coefficients, and flavon parameters. Each generation comprises three stages: (i) \emph{variation}, which creates an offspring set $Q$ from $P$ via crossover (exchanging parameter values between pairs of parents) and mutation (perturbing parameter values for exploration); (ii) \emph{non-dominated sorting}, which ranks all individuals in $P \cup Q$ by Pareto dominance; and (iii) \emph{selection}, which chooses the best $N$ individuals from $P \cup Q$ for the next generation.

Given the combined population of parents and offspring, $P \cup Q$, NSGA-III partitions this set into non-dominated fronts $\mathcal{F}_1, \dots, \mathcal{F}_k$, where $\mathcal{F}_1$ is the non-dominated (Pareto) front, $\mathcal{F}_2$ contains solutions dominated only by those in $\mathcal{F}_1$, and so on until all points are assigned.\footnote{Thus $k$ is not fixed.} More precisely, for each individual $\mathbf{x} \in P \cup Q$, we compute the dominance count $n_d(\mathbf{x}) = |\{\mathbf{y} : \mathbf{y} \prec \mathbf{x}\}|$ 
which record how many candidate solutions dominate $\mathbf{x}$. The first front consists of all individuals with $n_d(\mathbf{x}) = 0$, and subsequent fronts are constructed iteratively by removing previously assigned individuals. The new generation is formed during selection by sequentially collecting the candidates from the least to the most dominated fronts until reaching the target population size, $N$. When the next front would exceed $N$, candidates from the boundary front are selected to maximise diversity using reference directions, as explained next. Since the new generation is populated form the combined population of parents and offspring, $P \cup Q$, it might include parents as well as offspring, a feature called ``elitism'' wherein the best points are not lost through combination, but might be altered via mutation.

NSGA-III uses a reference-direction-based selection mechanism 
to identify a diverse set of candidates.
The key feature of NSGA-III is its use of a structured set of reference directions distributed across the objective space. For an $n_O$-dimensional problem with $H$ user-defined number of reference directions, these are built by placing reference points on the hyperplane $\sum_{i=1}^{n_O} \mathcal{L}_i = 1$ in the normalised objective space and drawing rays from the origin through each point. These directions, $r_j$ for $j = 1, \dots, H$, define the \emph{preference structure} that steers the population toward a diverse Pareto approximation.

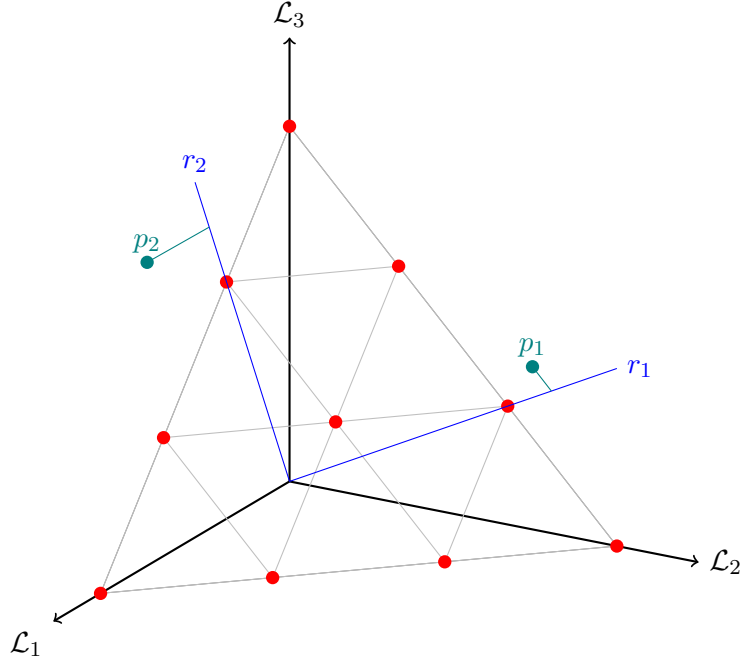
\begin{figure}[h]
    \centering
\tdplotsetmaincoords{70}{120}

\begin{tikzpicture}[
    scale=5,
    tdplot_main_coords,
    simplex/.style={very thin, gray},
    reference/.style={blue},
    points/.style={teal}
    ]
	\begin{scope}[thick]
		\draw[->] (0,0,0) -- (1.25,0,0) node[anchor=north east]{$\mathcal{L}_1$};
		\draw[->] (0,0,0) -- (0,1.25,0) node[anchor=west]{$\mathcal{L}_2$};
		\draw[->] (0,0,0) -- (0,0,1.25) node[anchor=south]{$\mathcal{L}_3$};
	\end{scope}

	\draw[simplex] (1,0,0) -- (0,1,0) -- (0,0,1) -- cycle;

	\begin{scope}[gray!50, thin]
		\foreach \x in {0,0.33333333333,...,1.0} {
				\draw (1-\x,0,\x) -- (0,1-\x,\x);
                \draw (1-\x,\x,0) -- (1-\x,0,\x);
                \draw (1-\x,\x,0) -- (0,\x,1-\x);
			}
	\end{scope}

	\foreach \x in {0,1,...,3} {
			\foreach \y in {0,1,...,3} {
					\foreach \z in {0,1,...,3} {
							\pgfmathtruncatemacro{\sumxyz}{\x + \y + \z}
							\ifnum\sumxyz=3 
								\coordinate (pt) at (\x/3, \y/3, \z/3);
								\fill[red] (pt) circle (0.5pt);
							\fi
						}
				}
		}

    \draw[reference] (0,0,0) -- (0,1,0.5) node[right]{$r_1$};
    \fill[points] (0.1, 0.8, 0.5) circle (0.5pt);
    \draw[points] (0.1, 0.8, 0.5) node[above]{$p_1$};
    \draw[points] (0.1, 0.8, 0.5) -- (0.0, 0.8, 0.4);

    \draw[reference] (0,0,0) -- (0.5,0,1) node[above]{$r_2$};
    \fill[points] (1.1, 0.2, 1.0) circle (0.5pt);
    \draw[points] (1.1, 0.2, 1.0) node[above]{$p_2$};
    \draw[points] (1.1, 0.2, 1.0) -- (0.425, 0.0, 0.85);

\end{tikzpicture}
    \caption{Illustration depicting the scheme for constructing the reference directions ($r_i$, as blue lines) from the reference points (red points) in NSGA-III, and how points in the constraint space ($p_i = (\mathcal{L}_1(x_i),\dots)$, as green) are assigned to a reference direction. Adapted from~\cite{deSouza:2025uxb}.}
    \label{fig:reference_directions}
\end{figure}

For a candidate solution $\mathbf{x}$ with objective vector $(\mathcal{L}_1(\mathbf{x}), \dots, \mathcal{L}_{n_O}(\mathbf{x}))$, we first apply a piecewise linear transformation to scale each objective to normalise it to the range $[0, 1]$:
\begin{equation}
    \mathcal{L}'_i(\mathbf{x}) = \frac{\mathcal{L}_i(\mathbf{x}) - \mathcal{L}^{\min}_i}{\mathcal{L}^{\max}_i - \mathcal{L}^{\min}_i}\,,
\end{equation}
where $\mathcal{L}^{\max}_i$ and $\mathcal{L}^{\min}_i$ represent the maximum and minimum values of the $i$-th objective within the current population. Each solution is then associated with its nearest reference direction via the perpendicular distance metric
\begin{equation}
    d_{\perp}(\mathbf{x}, \mathbf{r}_j) = \left\| \mathcal{L}'(\mathbf{x}) - \left( \mathcal{L}'(\mathbf{x}) \cdot \hat{\mathbf{r}}_j \right) \hat{\mathbf{r}}_j \right\| \,,
\end{equation}
where $\hat{r}_j$ is the unit vector along direction $r_j$. The reference direction with minimal perpendicular distance is assigned to the solution. In~\cref{fig:reference_directions} we schematically show in three dimensions how two points are assigned to a reference direction. The hyperplane $\sum_{i=1}^{n_O} \mathcal{L}_i = 1$ is partitioned into \emph{reference points}, which define the reference directions. These points are chosen at the start of the scan and used for all generations. We use the standard \emph{Das–Dennis} algorithm to allocate the reference points, populating the hyperplane with equally spaced points. 

Therefore, during selection, when the population would exceed $N$, individuals from the last admitted front $\mathcal{F}_l$ are chosen based on their association with reference directions. For each direction $r_j$, we select the solution in $\mathcal{F}_l$ with the smallest perpendicular distance to $r_j$. Directions with no associated solutions are filled first, reinforcing sparsely represented regions of the Pareto front, increasing solution diversity, and thus enhancing exploration. This mechanism balances \emph{convergence} (progress toward the Pareto front at the origin in our problem) and \emph{diversity} (coverage across objective combinations).

\subsection{Implementation Details}

The scan explores a parameter space analogous to that defined in Ref.~\cite{Nishimura:2020nre}, in which the 18 free FN charges are restricted to the range $|Q(X)| \leq 9$. In addition, we assume the 45 FN Wilson coefficients to be of order unity by imposing $0.3 < |C^X_{ij}| < 3$, and we vary the flavon parameters in two different regimes to set a hierarchy between flavons\footnote{This ensures truly FN explanations for flavour data. In early runs we found that many solutions would have both vevs near unity, leading to democratic textures that can only produce the data through tuned cancellations.}
\begin{itemize}
    \item $\Phi_u$ dominating: $|\epsilon_u| \in [0.3,0.1]$ and $|\epsilon_d|\in [0.07, 0.03|$
    \item $\Phi_d$ dominating: $|\epsilon_u| \in [0.3,0.1]$ and $|\epsilon_d|\in [0.07, 0.03|$ \ ,
\end{itemize}
and $\alpha \in [0,2\pi]$, resulting in a total of 66 free parameters.\footnote{To avoid topological complications associated with the identification $\alpha \sim \alpha + 2\pi$, we parametrise $\alpha$ in terms of two auxiliary variables, $\alpha_x$ and $\alpha_y$, both defined on $[-1,1]$, and reconstruct $\alpha$ via $\alpha = \arctan(\alpha_y/\alpha_x)$. Consequently, the total number of parameters sampled by the algorithm is 67.} The FN charges of the flavon fields are fixed to $1$, and the FN charge of the Higgs doublet is determined by requiring a tree-level top-quark Yukawa coupling, which is likewise taken to be of order unity.

In this study, although the parameter space includes both discrete and continuous variables, all parameters were encoded as continuous within the GA operators, with discrete parameters later rounded appropriately. This choice was made for computational practicality: rigorously distinguishing between discrete and continuous parameters incurred prohibitive overhead, making the scan infeasible. Consequently, simulated binary crossover and polynomial mutation were applied to all parameters for recombination and mutation, respectively. A more detailed investigation of tailored genetic operators for different parameter types is left for future work.

We implemented NSGA-III with \texttt{pymoo}~\cite{pymoo} python package. The scan is executed in multiple parallel runs with different random seeds to ensure broad coverage of the parameter space. Each run terminates when one of the following criteria is met:
\begin{itemize}
    \item  A maximum number of generations, set to one million, is reached.
    \item A target number of unique valid models, set to 100, is found.
    \item The best-found loss or the number of satisfied constraints ceases to improve over 500 generations.
\end{itemize}
Upon completion, the combined set of models is deduplicated based on their FN charge assignments, ensuring that the results represent qualitatively distinct model configurations. In case of duplicates, the configuration with the observables closes to their central value are kept, effectively biasing the search to better fits.

\section{Results\label{sec:results}}

We now present the outcome of our scan, which constitutes the first detailed and systematic investigation of neutrino-physics predictions arising from two-flavon FN models. The results reported here are an collation of several independent and parallel scans. This strategy was adopted to obtain a more exhaustive characterisation of the space of phenomenologically viable FN realisations.

During the initial exploratory phase, we observed that allowing both flavon vevs to vary freely within the interval $[0.01,1]$ predominantly produced solutions of the form $\epsilon_u \sim \epsilon_d \lesssim 1$. Although these solutions satisfied all imposed phenomenological constraints, they effectively yielded nearly democratic Yukawa matrices, with agreement driven primarily by numerical ``coincidences'' in the Wilson coefficients. Such configurations do not exhibit the characteristic FN mechanism, in which hierarchical Yukawa textures and the observed mixing angles and mass hierarchies are generated by systematic suppression due to flavon vevs.

This observation motivated us to restrict one of the flavon vevs to the range $[0.1,0.3]$ and the other to $[0.03,0.07]$. Under these conditions, the resulting models display genuinely FN-like behaviour, where flavour structures are predominantly governed by flavon-induced suppression factors.

Additionally, without further guidance, our first scans yield almost exclusively normal ordering solutions, models with large values for the flavon charges, or with the maximal relative mass difference near the imposed limit, at $20\%$. While all agreeing with the constraints, and in a sense as good as a FN model as any other, we felt that we should steer the search algorithm towards solutions with specific features. Therefore, we conducted extra scans with the following settings
\begin{itemize}
    \item Force Inverted Ordered: we explicately searched for Inverted Order models by demanding $m_1>m_3$ by adding an extra loss of the form $C(m_1/m_3,1,\infty)$.
    \item Small Maximal Flavon Exponent: we explicately searched for models with the smallest passible value of $n^X_{ij}$ by adding an extra loss of the form $C(\max{n^X_{ij}},0,M)$, where we varied $M=3,4,\dots$ to find the smallest $M$ yielding viable models.
    \item Smaller Maximal Relative Mass Difference: we performed extra runs with the maximal relative mass difference set to $10\%$ instead of the original $20\%$.
\end{itemize}

\begin{figure}[h]
    \centering
    \includegraphics[width=0.45\linewidth]{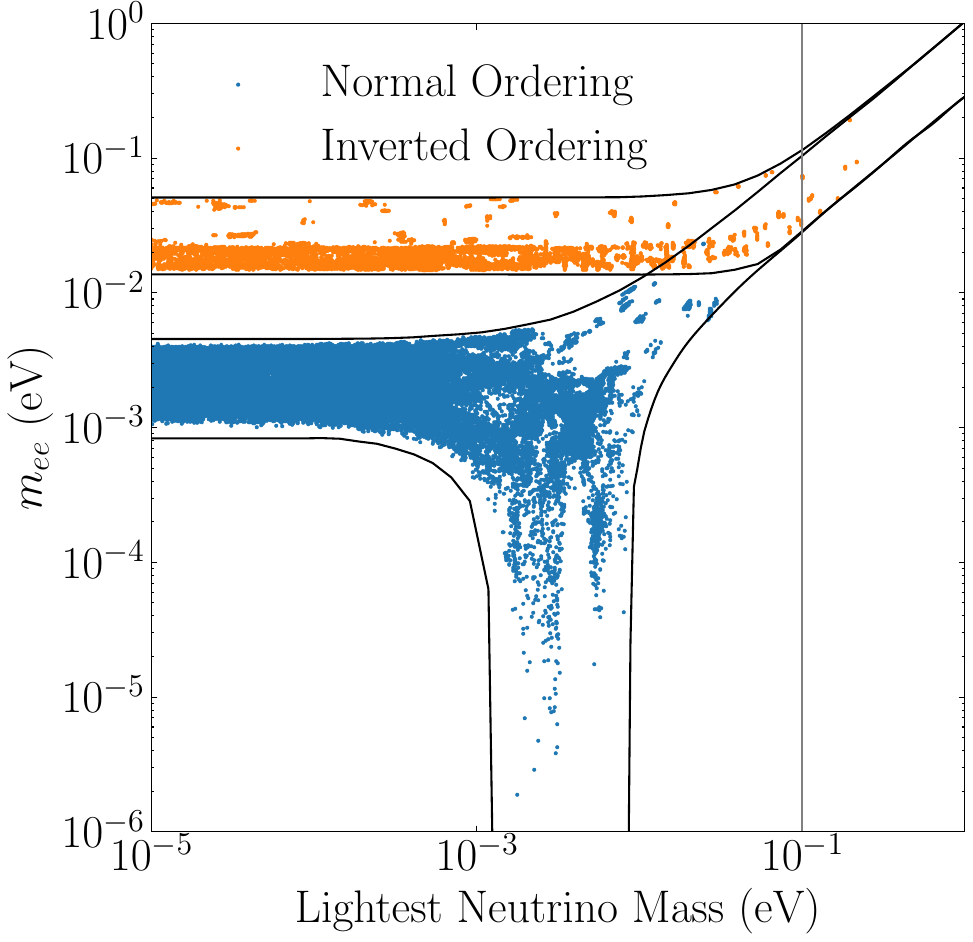}
    \caption{Discovered models with unique FN assignments in the $($lightest neutrino, $m_{ee})$ plane. Colour coded by Normal (Inverted) Ordering in Blue (Orange). Bounds represent the $3\sigma$ allowed region. Vertical line at $0.1$ eV roughly represents current cosmological upper bounds on the lightest neutrino mass.}
    \label{fig:all_results_io_vs_no}
\end{figure}

The aggregation of all the discovered models is presented in~\cref{fig:all_results_io_vs_no}. Here, we display the  models in the $($lightest neutrino mass, $m_{ee})$ plane, where $m_{ee}$ is the so-called effective neutrinoless double beta  mass, 
\begin{equation}
    m_{ee} = \left| \sum_i U_{ei}^2 m_i \right| \ ,
\end{equation}
where $U$ is the PMNS matrix. As it can be observed, the region within the \texttt{NuFit} bounds is well populated for the Normal Ordering models, whereas it proves more difficult to obtain Inverted Ordering predictions for higher values of $m_{ee}$. While tempting to retrieve a statistical interpretation from these figures, we need to reiterate that the way that NSGA-III produces results does not lead to statistically interpretable densities, either frequentist or Bayesian. NSGA-III, and similar evolutionary optimisation (and conceivably reinforcement learning as well) algorithms produce solutions that are \emph{easier} to obtain, producing a bias towards regions in the search space that provide quicker paths to the solution.

In total, we found $100\ 625$, of which $100\ 599$ are unique, i.e. our multiple independent scan strategy only produced $26$ duplicate models. This suggests that there many more viable models that could be found, potentially filling the scattered regions of~\cref{fig:all_results_io_vs_no}.

\begin{figure}[h]
    \centering
    \includegraphics[width=0.45\linewidth]{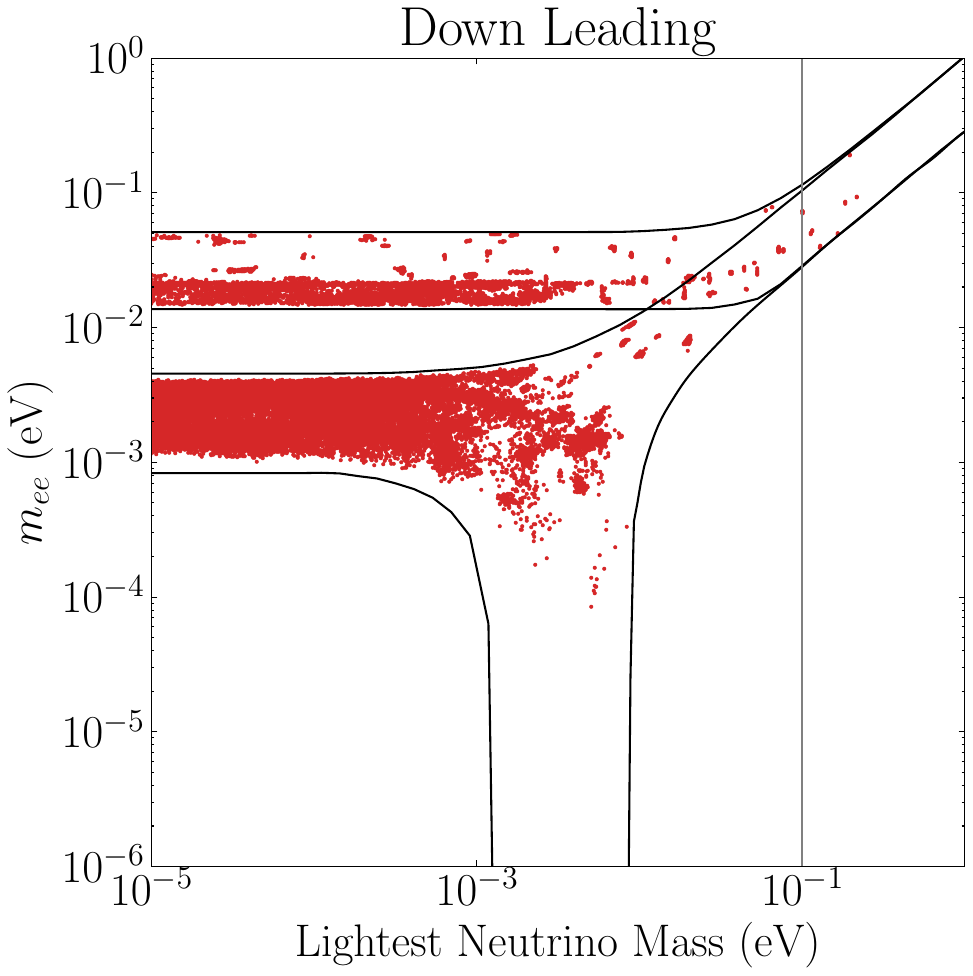}
    \includegraphics[width=0.45\linewidth]{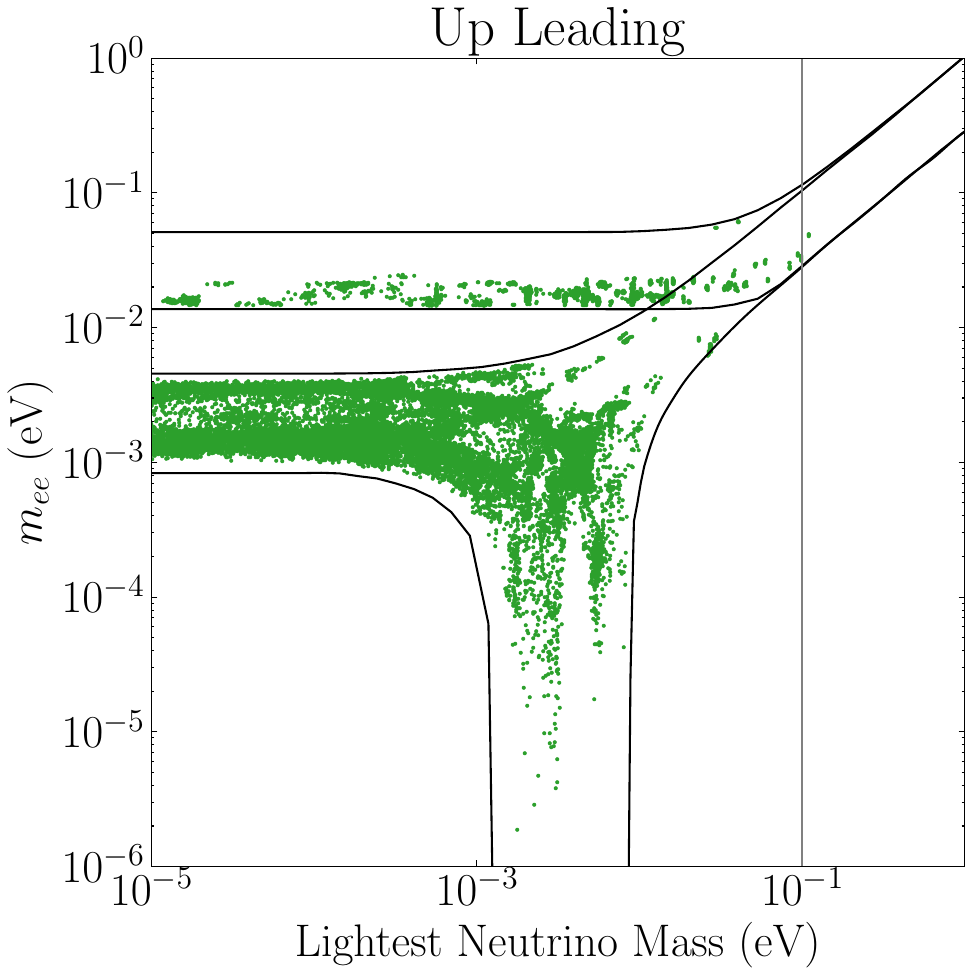}
    \caption{Discovered models with unique FN assignments in the $($lightest neutrino, $m_{ee})$ plane. Colour coded by Up-(Down-)leading Red (Green). Bounds represent the $3\sigma$ allowed region. Vertical line at $0.1$ eV roughly represents current cosmological upper bounds on the lightest neutrino mass.}
    \label{fig:all_results_up_vs_down}
\end{figure}

As discussed in the preceding sections, we impose a hierarchy between the flavon vevs, both taken to be sufficiently small such that the flavour structure is dominantly governed by the FN mechanism. For each of the scenarios, $\epsilon_u < \epsilon_d$ (Down Leading) and $\epsilon_u > \epsilon_d$ (Up Leading), we performed an equal number of parameter scans under identical settings. In~\cref{fig:all_results_up_vs_down}, we present the resulting models separated according to these two cases.

Although the overall picture is similar, several noteworthy differences emerge. The algorithm identifies Up Leading models with Normal Ordering that yield $m_{ee} < 10^{-4}$, a region that is difficult to realise within the Down Leading setup. Conversely, the algorithm finds multiple Inverted Ordering models with comparatively large values of $m_{ee}$ in the Down Leading case, while such solutions are almost absent in the Up Leading scenario. These high-$m_{ee}$ configurations are closest to the projected future sensitivities of neutrinoless double beta decay experiments and therefore indicate that a subset of FN-based models can be tested and potentially excluded by upcoming measurements. 

In both scenarios, only a small number of models with the lightest neutrino mass exceeding $0.1\,\text{eV}$ -- a value roughly corresponding to current cosmological upper bounds on the lightest neutrino mass -- are obtained.

A natural question to address is the degree of “minimality” of the models discovered through our methodology. In this context, “minimality” refers specifically to the values of the flavon vevs exponents, which determine the effective expansion order of the non-renormalizable FN operators. Smaller flavon vev exponents correspond to simpler tree-level structures in the UV completion that is integrated out at the FN scale $\Lambda$, whereas larger effective suppressions of the flavon vevs typically originate from more elaborate, non-minimal patterns of couplings in the UV between the SM fields and the heavy states that have been integrated out.  

\begin{figure}[h]
    \centering
    \includegraphics[width=0.55\linewidth]{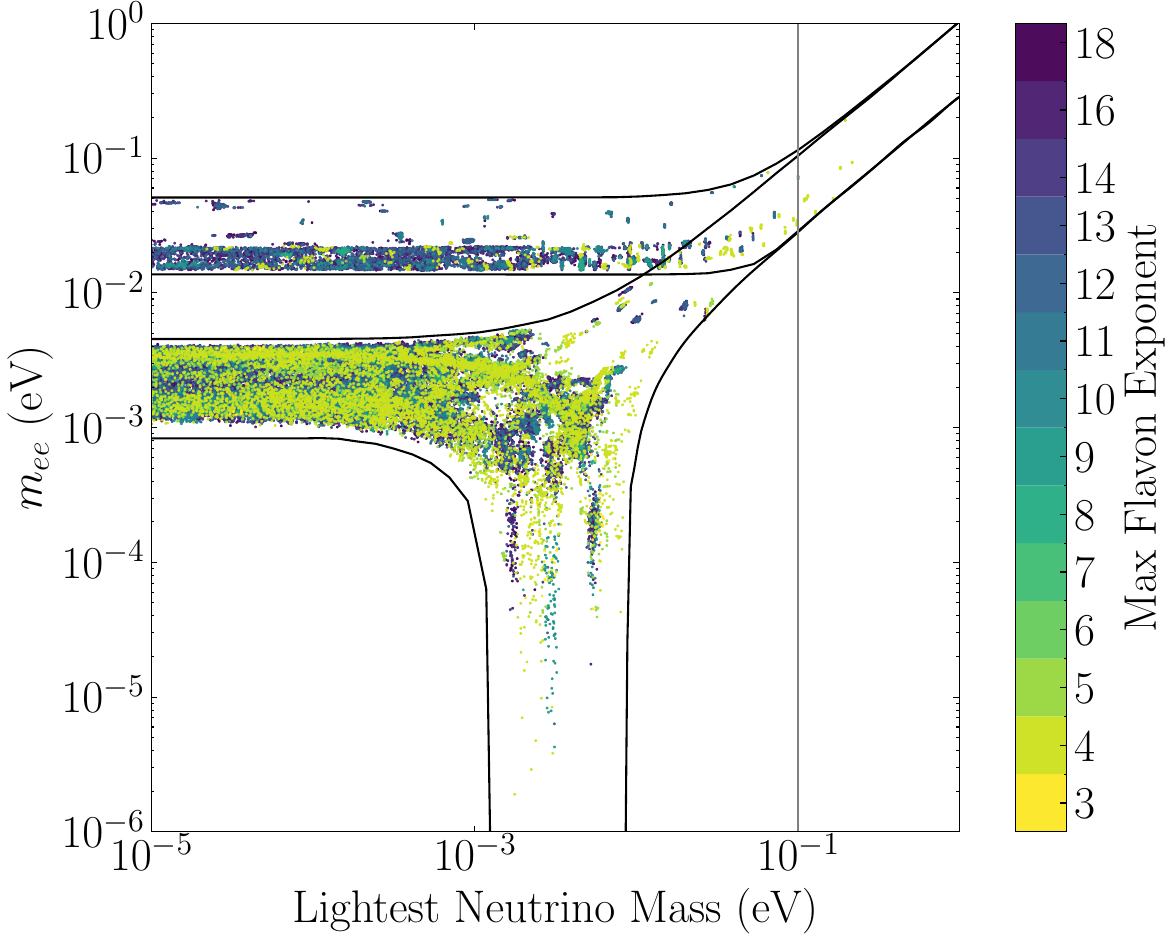}
    \caption{Discovered models with unique FN assignments in the $($lightest neutrino, $m_{ee})$ plane. Colour coded by maximual flavon exponent value. Bounds represent the $3\sigma$ allowed region. Vertical line at $0.1$ eV roughly represents current cosmological upper bounds on the lightest neutrino mass.}
    \label{fig:all_results_max_insertions}
\end{figure}

To quantitatively evaluate how minimal the FN constructions can be, we performed dedicated parameter scans in which we imposed upper bounds on the allowed flavon exponents. The global outcome of these scans, categorised by the maximal flavon exponent and colour-coded accordingly, is shown in~\cref{fig:all_results_max_insertions}, where points corresponding to smaller maximal exponents are plotted on top to facilitate their visual identification. From these results, we find that although very large maximal flavon exponents occur frequently, there exist viable FN models in which the largest exponent does not exceed three, thereby yielding comparatively “minimal” FN realisations.

A final question we address is how accurately our framework can reproduce the observed charged fermion masses. Our initial tolerance of $20\%$ was motivated by theoretical and experimental uncertainties, as well as by ambiguities in defining all mass parameters at a common renormalisation scale for this analysis. Although this constraint is significantly more stringent than that adopted in~\cite{Nishimura:2020nre}, it remains comparatively liberal.

\begin{figure}[h]
    \centering
    \includegraphics[width=0.55\linewidth]{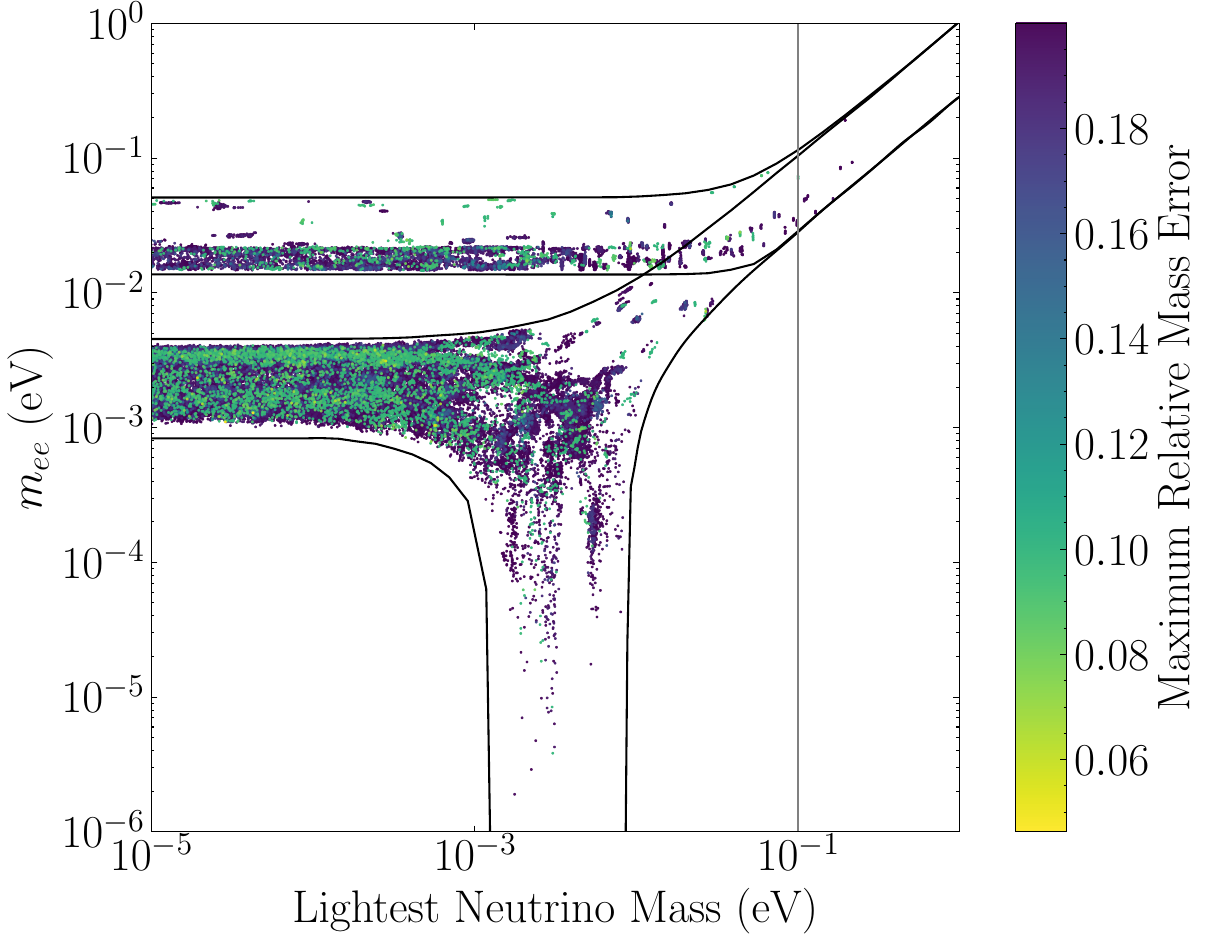}
    \caption{Discovered models with unique FN assignments in the $($lightest neutrino, $m_{ee})$ plane. Colour coded by charged fermions maximal relative mass error. Bounds represent the $3\sigma$ allowed region. Vertical line at $0.1$ eV roughly represents current cosmological upper bounds on the lightest neutrino mass.}
    \label{fig:all_results_max_mass_rel_error}
\end{figure}

While the primary aim of this work is not a detailed phenomenological fit, we performed a dedicated parameter scan targeting realisations with maximal relative mass deviations below $10\%$. The corresponding results are displayed in~\cref{fig:all_results_max_mass_rel_error}, where the maximal relative mass error for the charged fermions is colour-coded. We find that our methodology successfully identifies Froggatt–Nielsen realisations with relative mass discrepancies below $10\%$ for both inverted and normal neutrino mass ordering, and in some cases as small as $6\%$.

Given that our approach does not involve an explicit optimisation of continuous parameters with respect to a $\chi^2$ loss function, these findings constitute a striking demonstration of the effectiveness of our strategy for exploring mixed discrete–continuous parameter spaces, in which model structures and their continuous parameters are determined simultaneously.

Finally, we also examined the predictions for the PMNS parameters to assess whether the FN models discovered in this work yield any distinctive or characteristic signatures. As shown in~\cref{app:pmns_predictions}, no significant correlations or systematic patterns were observed in the resulting parameter space.

\section{Conclusions and Outlook\label{sec:conclusions}}

We have presented a systematic study of two-flavon FN models in the up and down type flavon scenario, where $\Phi_u$ and $\Phi_d$ couple independently to the up-type and down-type sectors. This class of models naturally accommodates CP violation through the relative phase between the two flavon vevs -- a feature absent in single-flavon setups, where all complex phases can be absorbed by field redefinitions.

To explore the high-dimensional, mixed discrete-continuous parameter space, we developed a scan framework based on the NSGA-III multi-objective genetic algorithm, treating each phenomenological constraint as a separate optimisation objective. Compared to a reinforcement learning approach developed in for a single flavon case in~\cite{Nishimura:2020nre}, our GA-based methodology simultaneously fits all 18 FN charges, 45 Wilson coefficients, and flavon parameters to both quark and lepton sectors without a separate optimisation step, and identifies viable models orders of magnitude faster under similar conditions.

We found over $100\,000$ unique phenomenologically viable models satisfying all 17 constraints -- CKM and PMNS angles and CP phases, charged fermion masses, neutrino squared-mass differences, and non-negative flavon exponents. The extremely low duplication rate across independent scans strongly suggests that the space of viable models is far from exhausted. Both Normal and Inverted Ordering solutions are realised, with Normal Ordering models populating the $(m_{\mathrm{lightest}}, m_{ee})$ plane consistently with current \texttt{NuFit} bounds, while Inverted Ordering models cluster at comparatively lower values of $m_{ee}$.

The relative hierarchy between the flavon vevs leads to qualitatively distinct predictions. Up Leading models readily achieve very small $m_{ee}$ for Normal Ordering, whereas Down Leading models more easily yield Inverted Ordering configurations with larger $m_{ee}$ values within reach of upcoming neutrinoless double beta decay experiments.

By imposing explicit upper bounds on the maximal flavon exponent, we demonstrated that viable models exist with the largest exponent not exceeding three, corresponding to comparatively simple UV completions. Dedicated scans with tightened mass constraints further identified models with maximal relative charged-fermion mass deviations below $10\%$, and in some cases as small as $6\%$, without any dedicated $\chi^2$ optimisation of continuous parameters.

Several directions for future work present themselves. On the methodological side, tailored genetic operators properly distinguishing discrete and continuous parameters may improve convergence and exploration, and a systematic comparison between multi-objective GA and RL approaches would establish best practices for AI-driven BSM model exploration. On the phenomenological side, post-processing the discovered models with dedicated $\chi^2$ fits, including renormalisation group running, would enable precision studies and direct comparison with forthcoming results from neutrino oscillation, neutrinoless double beta decay, and cosmological experiments. Finally, we call attention to the fact that this framework is applicable to any BSM scenario, being easily adapted and extendable to different model spaces that can be adequately encoded in a vector space, and we leave to future work its application to the exploration of more elaborate and open-ended model building challenges.

\section*{Acknowledgements}

MCR is supported by the STFC under Grant No. ST/T001011/1. SFK 
is funded by a Leverhulme Trust Emeritus Fellowship Grant.

\appendix

\section{CP Violation}

In this appendix, we review and synthesise the principal results concerning the origin of CP violation in Froggatt-Nielsen models, a subject that has been examined in~\cite{Kanemura:2007yy}. In particular, we explicitly demonstrate that, assuming real Wilson coefficients, one-flavon Froggatt-Nielsen models do not exhibit CP violation, whereas CP violation arises generically in models involving two flavon fields. 

\subsection{One Flavon\label{sec:cpv_single_flavon}}

In models with a single flavon field with real FN Wilson coefficients, the only possible source of CP violation is the a non-vanishing phase of the FN flavon vev. However, even this case is not a generic prediction, since a model with a $U(1)_{FN}$ flavour symmetry typically allows to perform a $U(1)_{FN}$ rotation to render the flavon vev real without the need to rephase the chiral matter fields. Nonetheless, for the sake of argument, we consider a FN set-up where the flavon vev carries a phase, $\langle \phi \rangle \propto e^{i\alpha}$. The flavon-induced phases entering the Yukawa couplings are then of the form $e^{i n_{ij} \alpha}$, where the integers $n_{ij}$ arise from the Froggatt–Nielsen exponents. As the exponents $n_{ij}$ are additive -- being given by sums of the relevant $U(1)_{FN}$ charges -- the associated phases are generically factorisable. Consequently, they can be absorbed by appropriate rephasings of the chiral matter fields. Therefore, a single flavon field does not, by itself, lead to physical CP-violating phases. We now demonstrate this explicitly.

Consider a single flavon field $\Phi$ with a vev parametrised as
\begin{equation}
	\langle\Phi\rangle = v_\Phi e^{i\alpha}.
\end{equation}
In the presence of a single flavon, the entries of the Yukawa matrices arise from higher-dimensional operators involving insertions of $(\Phi/\Lambda)^{n_{ij}}$. Schematically, one obtains
\begin{equation}
	Y_{ij}\sim c_{ij}\left(\frac{v_\Phi}{\Lambda}\right)^{n_{ij}} e^{i n_{ij}\alpha},
\end{equation}
where $c_{ij}$ are real coefficients of and $n_{ij}$ are non-negative integers determined by the $U(1)_{FN}$ charge-cancellation conditions. More concretely, we consider effective operators of the form
\begin{equation}
	\mathcal{L} \supset Y_{ij} \, H \, \bar \psi^i_L \, \psi^j_R \ ,
\end{equation}
for which the requirement of $U(1)_{FN}$ invariance implies
\begin{equation}
	n_{ij} \, Q(\Phi) = - Q(H) + Q(\psi^i_L) - Q(\psi^j_R) \ ,
\end{equation}
where $Q(X)$ denotes the $U(1)_{FN}$ charge of the field $X$. For definiteness, we assume that the charges can be normalised such that $Q(\Phi)=1$, which yields
\begin{equation}
	n_{ij} = - Q(H) + Q(\psi^i_L) - Q(\psi^j_R) \ ,
\end{equation}
and consequently the flavon-induced phases take the form
\begin{equation}
	n_{ij} \alpha = \alpha \big( - Q(H) + Q(\psi^i_L) - Q(\psi^j_R) \big) \ .
\end{equation}
This relation demonstrates that the phase structure of the Yukawa matrix is factorizable: the phases decompose into contributions associated with the left-handed and right-handed fields, respectively. As a result, all complex phases can be eliminated by appropriate field redefinitions, as we will discuss in the following.

We now consider the following rephasings of the chiral fermion fields
\begin{align}
	\psi^i_L &\;\to\; e^{i \lambda_i}\,\psi^i_L  \,,\\
	\psi^i_R &\;\to\; e^{i \rho_i}\,\psi^i_R \,,
\end{align}
which in turn induce the following transformation of the Yukawa phases:
\begin{equation}
	n_{ij}\,\alpha \;\to\; n_{ij}\,\alpha - \lambda_i + \rho_j \,.
\end{equation}
Consequently, the Yukawa phases can be eliminated if the condition
\begin{equation}
	n_{ij}\,\alpha \;=\; \lambda_i - \rho_j
	\;\;\Leftrightarrow\;\;
	\alpha\bigl(-Q(H) + Q(\psi^i_L) - Q(\psi^j_R)\bigr)
	\;=\; \lambda_i - \rho_j
\end{equation}
is satisfied. Without loss of generality, we now choose
\begin{align}
	\lambda_i &= \alpha\,l_i \,,\\
	\rho_j    &= \alpha\,r_j \,,
\end{align}
and impose
\begin{align}
	l_i &= Q(\Psi^i_L) \,,\\
	r_j &= Q(\Psi^j_R) + Q(H) \,.
\end{align}
With this choice, all Yukawa couplings become real, where all relations are to be understood modulo $2\pi$. Thus, there exists sufficient freedom in the chiral field phase redefinitions to remove the complex flavon–induced phases from the Yukawa sector.\footnote{This conclusion can equivalently be established by counting parametric degrees of freedom. The flavon–induced phases introduce six independent phase parameters, which coincides with the number of independent phases available from the chiral field rephasings. Although one might be concerned that making the up-type Yukawa couplings real could in turn generate complex phases in the down-type Yukawa couplings -- seemingly leaving only three remaining phase degrees of freedom to render the down-type Yukawas real -- the phases thus induced are not independent. As a result, the three remaining rephasing degrees of freedom suffice to ensure that both the up- and down-type Yukawa matrices can be simultaneously made real.}

\subsection{Two Flavon Models\label{sec:cpv_two_flavons}}

Since the single-flavon scenario does not furnish a source of CP violation, we now turn to models with two flavons. Exploiting the $U(1)_{FN}$ freedom, we may, without loss of generality, choose one of the flavon vevs to be real, which we take to be $\Phi_u$. For this exercise, we choose the parametrisation
\begin{equation}
    \epsilon=\frac{\langle\Phi_u\rangle}{\Lambda}\,, \qquad R=\frac{\langle\Phi_d\rangle}{\langle\Phi_u\rangle} = |R|\, e^{i\alpha}\,,
\end{equation}
which, for the purpose of this discussion, it is irrelevant whether $R>1$ or $R<1$. It follows immediately that, for real Wilson coefficients $c^{u,\nu}_{ij}$,
\begin{align}
	Y^u_{ij} &= c^u_{ij}\,\epsilon^{\frac{n^u_{ij}}{Q(\Phi_u)}} \quad \Rightarrow \quad \Im(Y^u_{ij}) = 0\,, \\
	Y^\nu_{ij} &= c^\nu_{ij}\,\epsilon^{\frac{n^\nu_{ij}}{Q(\Phi_u)}} \quad \Rightarrow \quad \Im(Y^\nu_{ij}) = 0 \, .
\end{align}

The reality of $Y^u$ does not, however, by itself preclude CP violation, since the down-type Yukawa matrix $Y^d$ may still retain sufficient non-trivial complex structure to generate a physical CP-violating phase. We now investigate this possibility. The entries of $Y^d$ are given by
\begin{equation}
	Y^d_{ij} = c^d_{ij}\,\epsilon^{\frac{n^d_{ij}}{Q(\Phi_d)}}\,R^{\frac{n^d_{ij}}{Q(\Phi_d)}} \,,
\end{equation}
from which we see that its complex phase structure is of the form
\begin{equation}
	Y^d_{ij} \propto e^{i\alpha \left(\frac{n^d_{ij}}{Q(\Phi_d)}\right)} \,.
\end{equation}

As before, we now fix $Q(\Phi_d) = 1$, so that
\begin{equation}
	Y^d_{ij} \propto e^{i \alpha n^d_{ij}} \,,
\end{equation}
which already closely resembles the single-flavon setup discussed in~\cref{sec:cpv_single_flavon}. Indeed, we can perform field redefinitions of the form
\begin{align}
	Q^i &\to e^{i\alpha\,l_i}\,Q^i\,, \\
	d^i &\to e^{i\alpha\,r_i}\,d^i \,,
\end{align}
and choose $l_i$ and $r_i$ such that the phases in $Y^d$ are removed. This requires
\begin{equation}
	- \alpha l_i + \alpha r_j = \alpha n^d_{ij} = Q(\bar Q_i) + Q(d_j) + Q(H) \,,
\end{equation}
with the solution
\begin{align}
	l_i &= Q(Q_i)\,, \\
	r_j &= Q(d_j) + Q(H) \,.
\end{align}

These transformations render $Y^d$ real, i.e.\ $\Im(Y^d_{ij}) = 0$, but they induce new phases in $Y^u$, which was previously real:
\begin{equation}
	Y^u_{ij} \;\to\; Y^u_{ij}\,e^{-i\alpha l_i} \quad \text{(no sum over $i$)} \,.
\end{equation}

We now assess whether these induced phases can be removed by rephasing the right-handed up-type quark fields. Consider
\begin{equation}
	u^i \to e^{i\alpha s_i}\,u^i \,,
\end{equation}
which implies
\begin{equation}
	Y^u_{ij} e^{-i\alpha l_i} \quad \text{(no sum)} \;\to\; Y^u_{ij}\,e^{-i\alpha l_i}\,e^{i\alpha s_j} \quad \text{(no sum)} \,.
\end{equation}
To make all entries of $Y^u$ real, we must satisfy the system of equations
\begin{equation}
	l_i - s_j = 0 \,,
\end{equation}
for all $i,j$, i.e.\ nine conditions for the three independent phases $s_j$, with the $l_i$ fixed by the $U(1)_{FN}$ charges. The only possible solution is
\begin{align}
	s_1 &= l_1\,, \\
	s_2 &= l_1\,, \\
	s_3 &= l_1\,, \\
	l_1 &= l_2 = l_3 \,,
\end{align}
which occurs only if all $Q(Q_i)$ are equal. This corresponds to a non generic, highly constrained choice of flavour charges. Consequently, for generic charge assignments, it is not possible to render both Yukawa matrices simultaneously real by field redefinitions alone, and the model will in general exhibit CP violation.

An alternative, and perhaps more elegant, argument can be formulated by noting that the basis in which FN models are typically expressed is itself a weak basis. Consequently, as is well known in the case of three fermion families, CP violation can be quantified by the Jarlskog invariant~\cite{Jarlskog:1985ht,Botella:1994cs}
\begin{equation}
    J \propto \det\bigl([Y_u Y_u^\dagger,\, Y_d Y_d^\dagger]\bigr) \,,
\end{equation}
or, equivalently,
\begin{equation}
    J \propto \Im\{\text{Tr}(H_u H_d H_u^2 H_d^2)\}\,,
\end{equation}
where $H_k = Y_k Y_k^\dagger$ with $k = u,d$. From this point of view, the presence of CP violation can be assessed by computing these invariants directly from the predicted Yukawa matrices. Although the full analytical expressions are too lengthy to be displayed here, one arrives at an analogous conclusion: the Jarlskog invariant vanishes only if the charges of the matter fields are chosen in a highly specific way to enforce this cancellation, in agreement with the results obtained previously from a different viewpoint.

\section{PMNS Parameter Predictions\label{app:pmns_predictions}}

\begin{figure}[h]
    \centering
    \includegraphics[width=0.9\linewidth]{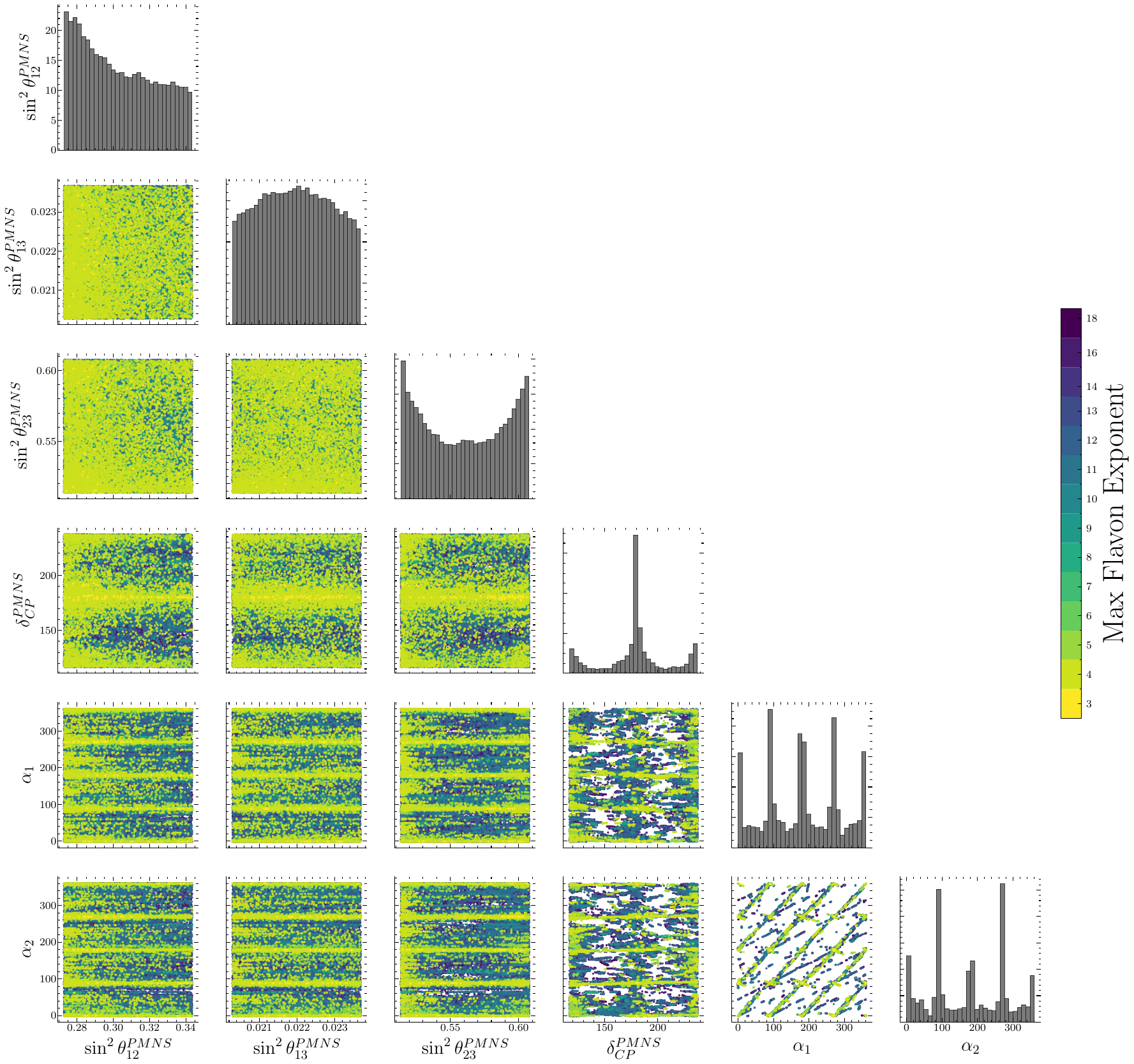}
    \caption{Predictions for the PMNS parameters, including the Majorana phases, for the unique FN models discovered. Colour coded by by maximal flavon exponent value.}
    \label{fig:pmns_predictions}
\end{figure}

In~\cref{fig:pmns_predictions} we present a corner plot that displays the predicted values for all PMNS parameters, including the Majorana phases. The scatter points are colour-coded according to the maximal flavon exponent, with smaller exponent values over-plotted to facilitate their visual identification. In these plots, we do not identify any discernible patterns or correlations that would yield clearly distinguishable predictions among the FN models identified in this work.

\bibliographystyle{unsrt}
\bibliography{references}

@article{Abel:2021rrj,
    author = "Abel, Steven and Constantin, Andrei and Harvey, Thomas R. and Lukas, Andre",
    title = "{Evolving Heterotic Gauge Backgrounds: Genetic Algorithms versus Reinforcement Learning}",
    eprint = "2110.14029",
    archivePrefix = "arXiv",
    primaryClass = "hep-th",
    doi = "10.1002/prop.202200034",
    journal = "Fortsch. Phys.",
    volume = "70",
    number = "5",
    pages = "2200034",
    year = "2022"
}

@article{Nishimura:2020nre,
    author = "Nishimura, Satsuki and Miyao, Coh and Otsuka, Hajime",
    title = "{Exploring the flavor structure of quarks and leptons with reinforcement learning}",
    eprint = "2304.14176",
    archivePrefix = "arXiv",
    primaryClass = "hep-ph",
    reportNumber = "KYUSHU-HET-257",
    doi = "10.1007/JHEP12(2023)021",
    journal = "JHEP",
    volume = "23",
    pages = "021",
    year = "2020"
}

@article{deb2013evolutionary,
  title={An evolutionary many-objective optimization algorithm using reference-point-based nondominated sorting approach, part I: solving problems with box constraints},
  author={Deb, Kalyanmoy and Jain, Himanshu},
  journal={IEEE transactions on evolutionary computation},
  volume={18},
  number={4},
  pages={577--601},
  year={2013},
  publisher={IEEE}
}

@article{jain2013evolutionary,
  title={An evolutionary many-objective optimization algorithm using reference-point based nondominated sorting approach, part II: Handling constraints and extending to an adaptive approach},
  author={Jain, Himanshu and Deb, Kalyanmoy},
  journal={IEEE Transactions on evolutionary computation},
  volume={18},
  number={4},
  pages={602--622},
  year={2013},
  publisher={IEEE}
}

@ARTICLE{pymoo,
    author={J. {Blank} and K. {Deb}},
    journal={IEEE Access},
    title={pymoo: Multi-Objective Optimization in Python},
    year={2020},
    volume={8},
    number={},
    pages={89497-89509},
}

@misc{hepmllivingreview,
    Author = "{HEP ML Community}",
    title = "{A Living Review of Machine Learning for Particle Physics}",
    url={https://iml-wg.github.io/HEPML-LivingReview/}
}

@article{deSouza:2022uhk,
    author = "de Souza, Fernando Abreu and Crispim Rom{\~a}o, Miguel and Castro, Nuno Filipe and Nikjoo, Mehraveh and Porod, Werner",
    title = "{Exploring parameter spaces with artificial intelligence and machine learning black-box optimization algorithms}",
    eprint = "2206.09223",
    archivePrefix = "arXiv",
    primaryClass = "hep-ph",
    doi = "10.1103/PhysRevD.107.035004",
    journal = "Phys. Rev. D",
    volume = "107",
    number = "3",
    pages = "035004",
    year = "2023"
}

@article{Romao:2024gjx,
    author = "Rom{\~a}o, Jorge Crispim and Crispim Rom{\~a}o, Miguel",
    title = "{Combining evolutionary strategies and novelty detection to go beyond the alignment limit of the Z3 3HDM}",
    eprint = "2402.07661",
    archivePrefix = "arXiv",
    primaryClass = "hep-ph",
    reportNumber = "IPPP/24/04, CFTP/24-002",
    doi = "10.1103/PhysRevD.109.095040",
    journal = "Phys. Rev. D",
    volume = "109",
    number = "9",
    pages = "095040",
    year = "2024"
}

@article{deSouza:2025uxb,
    author = "de Souza, Fernando Abreu and Castro, Nuno Filipe and Crispim Rom{\~a}o, Miguel and Porod, Werner",
    title = "{Exploring scotogenic parameter spaces and mapping uncharted dark matter phenomenology with multi-objective search algorithms}",
    eprint = "2505.08862",
    archivePrefix = "arXiv",
    primaryClass = "hep-ph",
    reportNumber = "IPPP/25/29",
    doi = "10.1007/JHEP10(2025)116",
    journal = "JHEP",
    volume = "10",
    pages = "116",
    year = "2025"
}

@article{deSouza:2025bpl,
    author = "de Souza, Fernando Abreu and Boto, Rafael and Crispim Rom{\~a}o, Miguel and Figueiredo, Pedro N. and Rom{\~a}o, Jorge C. and Silva, Jo{\~a}o P.",
    title = "{Unearthing large pseudoscalar Yukawa couplings with machine learning}",
    eprint = "2505.10625",
    archivePrefix = "arXiv",
    primaryClass = "hep-ph",
    reportNumber = "IPPP/25/28",
    doi = "10.1007/JHEP07(2025)268",
    journal = "JHEP",
    volume = "07",
    pages = "268",
    year = "2025"
}

@article{Boto:2025mmn,
    author = "Boto, Rafael and Rebelo, Tiago P. and Rom{\~a}o, Jorge C. and Silva, Jo{\~a}o P.",
    title = "{Machine Learning in the 2HDM2S model for Dark Matter}",
    eprint = "2509.01677",
    archivePrefix = "arXiv",
    primaryClass = "hep-ph",
    month = "9",
    year = "2025"
}

@article{Boto:2025ovp,
    author = "Boto, Rafael and Matos, Jo{\~a}o A. C. and Rom{\~a}o, Jorge C. and Silva, Jo{\~a}o P.",
    title = "{Surveying the complex three Higgs doublet model with Machine Learning}",
    eprint = "2510.02445",
    archivePrefix = "arXiv",
    primaryClass = "hep-ph",
    month = "10",
    year = "2025"
}

@article{Boto:2026gzj,
    author = {Boto, Rafael and Elyaouti, Karim and Fontes, Duarte and Gon{\c{c}}alves, Maria and M{\"u}hlleitner, Margarete and Rom{\~a}o, Jorge C. and Santos, Rui and Silva, Jo{\~a}o P.},
    title = "{Reassessing CP Violation in the C2HDM with Machine Learning}",
    eprint = "2601.15227",
    archivePrefix = "arXiv",
    primaryClass = "hep-ph",
    reportNumber = "KA-TP-03-2026, P3H-26-005",
    month = "1",
    year = "2026"
}

@article{Kanemura:2007yy,
    author = "Kanemura, Shinya and Matsuda, Koichi and Ota, Toshihiko and Petcov, Serguey and Shindou, Tetsuo and Takasugi, Eiichi and Tsumura, Koji",
    title = "{CP violation due to multi Froggatt-Nielsen fields}",
    eprint = "0704.0697",
    archivePrefix = "arXiv",
    primaryClass = "hep-ph",
    doi = "10.1140/epjc/s10052-007-0343-2",
    journal = "Eur. Phys. J. C",
    volume = "51",
    pages = "927--931",
    year = "2007"
}

@article{Jarlskog:1985ht,
    author = "Jarlskog, C.",
    title = "{Commutator of the Quark Mass Matrices in the Standard Electroweak Model and a Measure of Maximal CP Nonconservation}",
    reportNumber = "USIP-85-14",
    doi = "10.1103/PhysRevLett.55.1039",
    journal = "Phys. Rev. Lett.",
    volume = "55",
    pages = "1039",
    year = "1985"
}

@article{Botella:1994cs,
    author = "Botella, F. J. and Silva, Joao P.",
    title = "{Jarlskog - like invariants for theories with scalars and fermions}",
    eprint = "hep-ph/9411288",
    archivePrefix = "arXiv",
    reportNumber = "FTUV-94-68, IFIC-94-65",
    doi = "10.1103/PhysRevD.51.3870",
    journal = "Phys. Rev. D",
    volume = "51",
    pages = "3870--3875",
    year = "1995"
}

@article{ParticleDataGroup:2024cfk,
    author = "Navas, S. and others",
    collaboration = "Particle Data Group",
    title = "{Review of particle physics}",
    doi = "10.1103/PhysRevD.110.030001",
    journal = "Phys. Rev. D",
    volume = "110",
    number = "3",
    pages = "030001",
    year = "2024"
}

@article{Esteban:2024eli,
    author = "Esteban, Ivan and Gonzalez-Garcia, M. C. and Maltoni, Michele and Martinez-Soler, Ivan and Pinheiro, Jo{\~a}o Paulo and Schwetz, Thomas",
    title = "{NuFit-6.0: updated global analysis of three-flavor neutrino oscillations}",
    eprint = "2410.05380",
    archivePrefix = "arXiv",
    primaryClass = "hep-ph",
    reportNumber = "IFT-UAM/CSIC-24-140, YITP-SB-2024-24, IPPP/24/64, IPPP/24/64, IFT-UAM/CSIC-24-140, YITP-SB-2024-24",
    doi = "10.1007/JHEP12(2024)216",
    journal = "JHEP",
    volume = "12",
    pages = "216",
    year = "2024"
}

@article{Froggatt:1978nt,
    author = "Froggatt, C. D. and Nielsen, Holger Bech",
    title = "{Hierarchy of Quark Masses, Cabibbo Angles and CP Violation}",
    reportNumber = "CERN-TH-2519",
    doi = "10.1016/0550-3213(79)90316-X",
    journal = "Nucl. Phys. B",
    volume = "147",
    pages = "277--278",
    year = "1979"
}

@article{King:2003rf,
    author = "King, S. F. and Ross, Graham G.",
    title = "{Fermion masses and mixing angles from SU (3) family symmetry and unification}",
    eprint = "hep-ph/0307190",
    archivePrefix = "arXiv",
    reportNumber = "SHEP-03-14, CERN-TH-2003-147",
    doi = "10.1016/j.physletb.2003.09.027",
    journal = "Phys. Lett. B",
    volume = "574",
    pages = "239--252",
    year = "2003"
}

@article{Leurer:1993gy,
    author = "Leurer, Miriam and Nir, Yosef and Seiberg, Nathan",
    title = "{Mass matrix models: The Sequel}",
    eprint = "hep-ph/9310320",
    archivePrefix = "arXiv",
    reportNumber = "RU-93-43, WIS-93-93-PH",
    doi = "10.1016/0550-3213(94)90074-4",
    journal = "Nucl. Phys. B",
    volume = "420",
    pages = "468--504",
    year = "1994"
}

@article{Altmannshofer:2024hmr,
    author = "Altmannshofer, Wolfgang and Greljo, Admir",
    title = "{Recent Progress in Flavor Model Building}",
    eprint = "2412.04549",
    archivePrefix = "arXiv",
    primaryClass = "hep-ph",
    doi = "10.1146/annurev-nucl-121423-100950",
    journal = "Ann. Rev. Nucl. Part. Sci.",
    volume = "75",
    number = "1",
    pages = "201--322",
    year = "2025"
}

@article{Smolkovic:2019jow,
    author = "Smolkovi{\v{c}}, Aleks and Tammaro, Michele and Zupan, Jure",
    title = "{Anomaly free Froggatt-Nielsen models of flavor}",
    eprint = "1907.10063",
    archivePrefix = "arXiv",
    primaryClass = "hep-ph",
    doi = "10.1007/JHEP10(2019)188",
    journal = "JHEP",
    volume = "10",
    pages = "188",
    year = "2019",
    note = "[Erratum: JHEP 02, 033 (2022)]"
}

@article{Ibe:2024cvi,
    author = "Ibe, Masahiro and Shirai, Satoshi and Watanabe, Keiichi",
    title = "{Comprehensive Bayesian exploration of Froggatt-Nielsen mechanism}",
    eprint = "2412.19484",
    archivePrefix = "arXiv",
    primaryClass = "hep-ph",
    reportNumber = "IPMU24-0047",
    doi = "10.1007/JHEP03(2025)150",
    journal = "JHEP",
    volume = "03",
    pages = "150",
    year = "2025"
}

@article{Cornella:2023zme,
    author = "Cornella, Claudia and Curtin, David and Neil, Ethan T. and Thompson, Jedidiah O.",
    title = "{Mapping and probing Froggatt-Nielsen solutions to the quark flavor puzzle}",
    eprint = "2306.08026",
    archivePrefix = "arXiv",
    primaryClass = "hep-ph",
    reportNumber = "MITP-23-026",
    doi = "10.1103/PhysRevD.111.015042",
    journal = "Phys. Rev. D",
    volume = "111",
    number = "1",
    pages = "015042",
    year = "2025"
}

@article{Cornella:2024jaw,
    author = "Cornella, Claudia and Curtin, David and Krnjaic, Gordan and Mellors, Micah",
    title = "{Testing the Froggatt-Nielsen mechanism with lepton flavor and number violating processes}",
    eprint = "2501.00629",
    archivePrefix = "arXiv",
    primaryClass = "hep-ph",
    reportNumber = "FERMILAB-PUB-25-0010-T, CERN-TH-2024-215",
    doi = "10.1103/c2ws-hx4h",
    journal = "Phys. Rev. D",
    volume = "112",
    number = "11",
    pages = "115010",
    year = "2025"
}

@article{Cranmer:2019eaq,
    author = "Cranmer, Kyle and Brehmer, Johann and Louppe, Gilles",
    title = "{The frontier of simulation-based inference}",
    eprint = "1911.01429",
    archivePrefix = "arXiv",
    primaryClass = "stat.ML",
    doi = "10.1073/pnas.1912789117",
    journal = "Proc. Nat. Acad. Sci.",
    volume = "117",
    number = "48",
    pages = "30055--30062",
    year = "2020"
}

@article{Halverson:2019tkf,
    author = "Halverson, James and Nelson, Brent and Ruehle, Fabian",
    title = "{Branes with Brains: Exploring String Vacua with Deep Reinforcement Learning}",
    eprint = "1903.11616",
    archivePrefix = "arXiv",
    primaryClass = "hep-th",
    doi = "10.1007/JHEP06(2019)003",
    journal = "JHEP",
    volume = "06",
    pages = "003",
    year = "2019"
}

@article{Kawai:2024pws,
    author = "Kawai, Shinsuke and Okada, Nobuchika",
    title = "{Truth, beauty, and goodness in grand unification: A machine learning approach}",
    eprint = "2411.06718",
    archivePrefix = "arXiv",
    primaryClass = "hep-ph",
    doi = "10.1016/j.physletb.2024.139221",
    journal = "Phys. Lett. B",
    volume = "860",
    pages = "139221",
    year = "2025"
}

@article{Harvey:2021oue,
    author = "Harvey, T. R. and Lukas, A.",
    title = "{Quark Mass Models and Reinforcement Learning}",
    eprint = "2103.04759",
    archivePrefix = "arXiv",
    primaryClass = "hep-th",
    doi = "10.1007/JHEP08(2021)161",
    journal = "JHEP",
    volume = "08",
    pages = "161",
    year = "2021"
}

@inproceedings{Cole:2021nnt,
    author = "Cole, Alex and Krippendorf, Sven and Schachner, Andreas and Shiu, Gary",
    title = "{Probing the Structure of String Theory Vacua with Genetic Algorithms and Reinforcement Learning}",
    booktitle = "{35th Conference on Neural Information Processing Systems}",
    eprint = "2111.11466",
    archivePrefix = "arXiv",
    primaryClass = "hep-th",
    month = "11",
    year = "2021"
}

@article{Matchev:2024ash,
    author = "Matchev, Konstantin T. and Matcheva, Katia and Ramond, Pierre and Verner, Sarunas",
    title = "{Exploring the truth and beauty of theory landscapes with machine learning}",
    eprint = "2401.11513",
    archivePrefix = "arXiv",
    primaryClass = "hep-ph",
    doi = "10.1016/j.physletb.2024.138941",
    journal = "Phys. Lett. B",
    volume = "856",
    pages = "138941",
    year = "2024"
}

@article{Hammad:2022wpq,
    author = "Hammad, A. and Park, Myeonghun and Ramos, Raymundo and Saha, Pankaj",
    title = "{Exploration of parameter spaces assisted by machine learning}",
    eprint = "2207.09959",
    archivePrefix = "arXiv",
    primaryClass = "hep-ph",
    doi = "10.1016/j.cpc.2023.108902",
    journal = "Comput. Phys. Commun.",
    volume = "293",
    pages = "108902",
    year = "2023"
}

@article{Goodsell:2022beo,
    author = "Goodsell, Mark D. and Joury, Ari",
    title = "{Active learning BSM parameter spaces}",
    eprint = "2204.13950",
    archivePrefix = "arXiv",
    primaryClass = "hep-ph",
    doi = "10.1140/epjc/s10052-023-11368-3",
    journal = "Eur. Phys. J. C",
    volume = "83",
    number = "4",
    pages = "268",
    year = "2023"
}

@article{Baretz:2025zsv,
    author = "Baretz, Jason Benjamin and Fieg, Max and Ganesh, Vijay and Ghosh, Aishik and Knapp-Perez, V. and Rudolph, Jake and Whiteson, Daniel",
    title = "{Towards AI-assisted Neutrino Flavor Theory Design}",
    eprint = "2506.08080",
    archivePrefix = "arXiv",
    primaryClass = "hep-ph",
    reportNumber = "UCI-TR-2025-05",
    doi = "10.1038/s42005-026-02627-2",
    month = "6",
    year = "2025"
}

@article{deSouza:2026jww,
    author = "de Souza, Fernando Abreu and Castro, Nuno Filipe and Crispim Rom{\~a}o, Miguel and Goodsell, Mark D. and Ibrahimov, Farid and Porod, Werner",
    title = "{BSMArt 2: simpler and faster parameter space scans}",
    eprint = "2606.05410",
    archivePrefix = "arXiv",
    primaryClass = "hep-ph",
    reportNumber = "IPPP/26/43",
    month = "6",
    year = "2026"
}

\end{document}